\documentclass[aps,preprint,showpacs,superscriptaddress,groupedaddress]{revtex4}  
\usepackage[dvips]{graphicx}
\usepackage{dcolumn}   
\usepackage{bm}        
\usepackage{amssymb}   
\usepackage[utf8]{inputenc}
\usepackage{mathrsfs}
\usepackage{amsmath}
\usepackage{hyperref}
\usepackage{subfigure}
\usepackage{color}

\begin{document}

\title{Scattering of the $\varphi^8$ kinks with power-law asymptotics}

\author{Ekaterina Belendryasova}
\affiliation{National Research Nuclear University MEPhI (Moscow Engineering Physics Institute), Kashirskoe shosse 31, Moscow 115409, Russia}
\author{Vakhid A. Gani}
\email{vagani@mephi.ru}
\affiliation{National Research Nuclear University MEPhI (Moscow Engineering Physics Institute), Kashirskoe shosse 31, Moscow 115409, Russia}
\affiliation{Theory Department, National Research Center Kurchatov Institute, Institute for Theoretical and Experimental Physics, Bolshaya Cheremushkinskaya st.\ 25, Moscow 117218, Russia}

\begin{abstract}
We study the scattering of the $\varphi^8$ kinks off each other, namely, we consider those $\varphi^8$ kinks that have power-law asymptotics. The slow power-law fall-off leads to a long-range interaction between the kink and the antikink. We investigate how the scattering scenarios depend on the initial velocities of the colliding kinks. In particular, we observe the `escape windows' --- the escape of the kinks after two or more collisions, explained by the resonant energy exchange between the translational and vibrational modes. In order to elucidate this phenomenon, we also analyze the excitation spectra of a solitary kink and of a composite kink+antikink configuration.
\end{abstract}



\maketitle

\section{Introduction}
\label{sec:Introduction}

Topological solitons (kinks) in $(1+1)$-dimensional non-linear field-theoretical models are of growing interest for theoretical physics from high energy physics and cosmology to condensed matter \cite{vilenkin01,manton01,aek01}. This interest can be motivated by a number of reasons. First, the dynamics of certain physical systems can be described by $(1+1)$-dimensional models. Second, the structure and dynamics of many objects in $(2+1)$ and $(3+1)$ space-time dimensions can be modeled by the $(1+1)$-dimensional theory. For example, a three- or two-dimensional domain wall generated by a real scalar field looks like a one-dimensional topological field configuration (a soliton, or a kink) if viewed from the perpendicular direction. Topological defects in $(1+1)$ dimensions exist also in more complex models with two or more fields. In \cite{Lensky.JETP.2001,Kurochkin.CMMP.2004} the authors investigated the soliton-like configurations in the model with one real and one complex scalar fields. In \cite{Bazeia.PLA.2013,Bazeia.AHEP.2013,Alonso.PhysicaD.2018,Alonso.PRD.2018,Katsura.PRD.2014,Correa.PLB.2014} the kink-like structures were studied in models with two interacting real scalar fields, while in \cite{Bazeia.EPJC.2014} some interesting results for several real scalar fields were obtained. The authors of \cite{GaLiRa,GaLiRaconf} have found an exact analytic solution that describes a domain wall with a localized configuration of the scalar triplet on it. A method for calculating the bubble profile in a bounce solution for a multi-field potential with a false vacuum was elaborated in \cite{Akula.EPJC.2016}. In \cite{Ashcroft.JPA.2016} kinks collisions in a two-field model with multiple vacua were studied.

Interactions of kinks with each other and with spatial defects (impurities) are of great importance and have attracted the attention of physicists and mathematicians for a long time. The history of the subject is rather old and vast, see, e.g., the review \cite{aek01}. The first studies of the kink-antikink collisions date back to the 1970s and 1980s. For example, the author of Ref.~\cite{Kudryavtsev1975} observed the formation of a large-amplitude bion in the collisions of the kink and antikink of the $\varphi^4$ model at the initial velocity $v_\mathrm{in}=0.1$ (in units of the speed of light). The bion is a long-living bound state of the kink and antikink, and its formation had been a surprise, the common expectation being that the kink and antikink would annihilate, with all their energy being emitted in the form of small-amplitude waves. Other works of that time established that the capture and the bion formation occur at initial velocities below a certain critical value, $v_\mathrm{in}<v_\mathrm{cr}$, while at $v_\mathrm{in}>v_\mathrm{cr}$ the kinks would bounce off each other and escape. Besides that, at $v_\mathrm{in}<v_\mathrm{cr}$ a new interesting phenomenon was observed --- the so-called `escape windows', see, e.g., Refs.~\cite{Campbell.phi4.1983,Anninos.PRD.1991,Goodman.PRL.2007}. This was understood to be a consequence of the resonant energy exchange between the translational and the vibrational modes of the kinks.

Later the resonance phenomena have been found and investigated in other models, such as the modified sine-Gordon \cite{Peyrard.msG.1983} and the double sine-Gordon \cite{Campbell.1986,Campbell.dsG.1986,Malomed.PLA.1987,Malomed.PLA.1989,GaKuPRE,Gani.arXiv.2017.dsg}. Moreover, it was shown that the resonance energy exchange between the translational and the vibrational modes is possible even if the kink's excitation spectrum does not contain any vibrational modes \cite{GaKuPRE,dorey}. In this case the energy was accumulated in the vibrational mode of the composite kink+antikink configuration. Note that in some processes  the resonance frequency was different from the frequency of the vibrational mode of the kink(s).

Today, the study of the properties of topological defects is a vast and very fast developing area. Interactions of kinks with each other and with impurities have been studied \cite{oliveira01,oliveira02,Bazeia.arXiv.2017.sinh,Bazeia.arXiv.2017.sinh.conf,Malomed.JPA.1992,krusch01,saad01,saad02,saad03,rad1,rad2,Simas.JHEP.2016,Ahlqvist:2014uha,Mohammadi.PTP.2012,Mohammadi.PTEP.2014}; see also results for branes \cite{Bazeia.IJMPA.2017}, bubble-like structures \cite{Gonzalez.PRE.2017,Gonzalez.JCAP.2018}, interactions of breathers \cite{Popov.CMMP.2014}, and so on \cite{Bazeia.EPL.2015,Brihaye.PRD.2015,Gonzalez.PLA.2017,Lopez.NPA.2017,Katsimiga.PLB.2015,Popov.CMMP.2013,Gumerov.CMMP.2014}. Many interesting and important results have been obtained in the (1+1)-dimensional models with polynomial potentials: $\varphi^4$, $\varphi^6$, $\varphi^8$, and so on \cite{lohe,GaKuLi,MGSDJ,khare,GaLeLi,GaLeLiconf,Radomskiy,Weigel.conf.2017,Weigel.PLB.2017,Dorey.JHEP.2017,Bazeia.PRD.2006,He.arXiv.2016,Moradi.CNSNS.2017,Belendryasova.conf.2017,HOFT_chapter}. An impressive progress has been achieved in the studies of Q-balls \cite{Schweitzer.PRD.2012.1,Schweitzer.PRD.2012.2,Schweitzer.NPA.2016,Nugaev.PRD.2013,Bazeia.EPJC.2016,Bazeia.PLB.2016,Bazeia.PLB.2017,Dzhunushaliev.PRD.2016}, embedded topological defects, e.g., a skyrmion on a domain wall \cite{jennings,nitta5}, a Q-lump on a domain wall \cite{blyankinshtein}, fermionic states on a domain wall \cite{Campanelli.IJMPD.2004,GaKsKu01,GaKsKu02,Bazeia.EPJC.2017,Bazeia.fermion.2017,Bazeia.PLB.2018}, etc.~\cite{nitta1,nitta2,nitta3,nitta4,Loginov.YadFiz.2011,Bazeia.AnnPhys.2018}. Topologically non-trivial field configurations could be responsible for a variety of phenomena observed in the early Universe \cite{GaKiRu,GaKiRu.conf}. Notice also results on interaction solutions in $(2+1)$ dimensional non-linear equations, see \cite{Ma.CMA.2018} and references therein.

Apart from numerically solving the equations of motion, various approximate methods are widely used for modeling of the kink-antikink interactions. In particular, the collective coordinate approximation \cite{GaKuLi,Weigel.cc.2014,Weigel.cc.2016,Dorey.PLB.2018,Demirkaya.cc.2017,Baron.cc.2014,Javidan.cc.2010,Christov.cc.2008,GaKu.SuSy.2001} treats the kink+antikink configuration as a system with one or several degrees of freedom. For instance, the distance between kink and antikink can be considered as a single (translational) degree of freedom. Modifications of the collective coordinate method, which include additional degrees of freedom (e.g., vibrational ones), have been elaborated, see, e.g., \cite{Weigel.cc.2014,Weigel.cc.2016,Demirkaya.cc.2017}.

Another approximation is the Manton's method \cite[Ch.~5]{manton01}, \cite{perring62,rajaram77,Manton.npb.1979,KKS.PRE.2004}. This method estimates the force between the kink and (anti)kink at large separations using the asymptotics of the kinks. Note that at the moment the applicability of the method is well-proven only for kinks and solitons with exponential asymptotics.

Certain $(1+1)$ dimensional field-theoretical models with polynomial potentials have kinks with power-law asymptotics (at one or both spatial infinities) \cite{lohe,khare,Radomskiy,Bazeia.JPC.2018} (see also \cite{Guerrero.PRE.1997,Guerrero.PLA.1998,Guerrero.Phys_A.1998} for some other results on long-range interaction of kinks). Interactions of such kinks are not well investigated, therefore their detailed study is of current interest \cite{Gomes.PRD.2012}. Due to the power-law tails, the kinks `feel' and perturb each other at much larger distances compared to the case of exponential tails. This our paper presents the first systematic study of the interaction of kinks with power-law tails.

In this paper we study the scattering of the kink and the antikink of the $\varphi^8$ model \cite{lohe,khare,GaLeLi,GaLeLiconf,Radomskiy,Belendryasova.conf.2017}. This model is employed, e.g., to model massless mesons with self-interaction \cite{lohe}, and to describe successive phase transitions \cite{khare}. Each kink (antikink) of the $\varphi^8$ model has power-law asymptotics at one spatial infinity and exponential asymptotics at the other. In our numerical simulation, we start from the initial configuration in the form of the kink and the antikink, which are separated by a large distance and are turned with their power-law tails towards each other. The initial velocities of the kinks are equal in the laboratory frame of reference.

For the used ansatz we found that the kink and antikink repel each other. There are two critical values of the initial velocity, $v^{(1)}_\mathrm{cr}$ and $v^{(2)}_\mathrm{cr}$, which separate three different regimes of the kink-antikink scattering. At the initial velocities $v_\mathrm{in}<v^{(1)}_\mathrm{cr}$ kinks do not collide. At $v^{(1)}_\mathrm{cr}<v_\mathrm{in}<v^{(2)}_\mathrm{cr}$ the incident kinks become trapped, they form a bound state (a bion), but there is also a pattern of `escape windows', within which the kinks are able to escape. For $v_\mathrm{in}>v^{(2)}_\mathrm{cr}$ the two incident kinks always escape after the collision.

In the range $v^{(1)}_\mathrm{cr}<v_\mathrm{in}<v^{(2)}_\mathrm{cr}$ we found a complicated pattern of `escape windows' --- narrow intervals of the initial velocity that allow the kink and antikink escape to infinities after two, three, or more collisions. We also find the connection between the escape windows and the frequencies of the vibrational modes of the composite kink+antikink configuration. We analyze the small oscillations of the field between the collisions, and investigate the excitation spectrum of the kink+antikink configuration.

We also investigate the oscillations in the situation when the field homogeneously deviates from the vacuum value in a large space domain. Such oscillations could contribute in the kink-antikink interaction, because due to the power-law tails the kinks substantially affect the values of the field even at large distances.

Our paper is organized as follows. In Section \ref{sec:Model} we briefly discuss field-theoretical models in $(1+1)$ dimensional space-time that have topologically non-trivial solutions --- the kinks. We also introduce the $\varphi^8$ model, write out its kinks, and discuss their properties. In Section \ref{sec:Interactions} we present a detailed numerical study of the kink-antikink scattering. Section \ref{sec:Spectra} presents the analyses of the excitation spectra of an isolated kink, and of the composite kink+antikink configuration. The results of the analysis of small oscillations of the field at the collision point are also given in this Section. In Section \ref{sec:Vacua} we deal with spatial homogeneous oscillations of the field near a vacuum value. Finally, in Section \ref{sec:Conclusion} we give an outlook and the conclusion.

\section{General properties of models with polynomial potentials. Kinks of the $\boldsymbol\varphi^8$ model}
\label{sec:Model}

Consider a field-theoretical model with a real scalar field in $(1+1)$-dimensional space-time with its dynamics defined by the Lagrangian
\begin{equation}
\label{eq:lagrangian}
\mathcal{L} = \frac{1}{2}\left(\frac{\partial \varphi}{\partial t}\right)^2-\frac{1}{2}\left(\frac{\partial \varphi}{\partial x}\right)^2-V(\varphi),
\end{equation}
where $V(\varphi)$ is the potential, which defines the self-interaction of the field $\varphi$. The Lagrangian \eqref{eq:lagrangian} yields the equation of motion for the field $\varphi(x,t)$:
\begin{equation}
\label{eq:eqmo}
\frac{\partial^2\varphi}{\partial t^2} - \frac{\partial^2\varphi}{\partial x^2} + \frac{dV}{d\varphi} = 0.
\end{equation}
We are interested in the models with non-negative potentials having two or more degenerate minima (vacua of the model) $\varphi_1^{{\scriptsize \mbox{(vac)}}}$, $\varphi_2^{{\scriptsize \mbox{(vac)}}}$, and so on, with $V(\varphi_1^{{\scriptsize \mbox{(vac)}}})=V(\varphi_2^{{\scriptsize \mbox{(vac)}}})=...=0$. The energy functional for the field $\varphi$ is
\begin{equation}
\label{eq:energy}
E[\varphi] = \int_{-\infty}^{\infty}\left[\frac{1}{2} \left( \frac{\partial\varphi}{\partial t} \right)^2+\frac{1}{2} \left( \frac{\partial\varphi}{\partial x} \right) ^2+V(\varphi)\right]dx.
\end{equation}

In the static case the equation of motion \eqref{eq:eqmo} takes the form
\begin{equation}
\label{eq:steqmo}
\frac{d^2\varphi}{dx^2} = \frac{dV}{d\varphi},
\end{equation}
and the energy of a static configuration can be found as
\begin{equation}
\label{eq:stenergy}
E[\varphi]=\int_{-\infty}^{\infty}\left[\frac{1}{2} \left(\frac{d\varphi}{dx} \right)^2+V(\varphi)\right]dx.
\end{equation}
In order for this integral to be convergent, i.e.\ for the energy of a configuration to be finite, it is necessary that the field tends to the vacuum values at $x\to\pm\infty$, i.e.
\begin{equation}
\label{eq:two}
\lim_{x\to-\infty}\varphi(x) = \varphi_i^{{\mathrm{(vac)}}},
\quad
\lim_{x\to+\infty}\varphi(x) = \varphi_j^{{\mathrm{(vac)}}}.
\end{equation}
From Eq.~\eqref{eq:steqmo} we can easily obtain a first order differential equation for the function $\varphi(x)$:
\begin{equation}
\frac{d\varphi}{dx}=\pm\sqrt{2V}.
\label{eq:BPS}
\end{equation}
If the potential $V(\varphi)$ has two or more degenerate minima, the set of all static configurations with finite energy can be split into disjoint equivalence classes --- topological sectors --- according to their limiting behavior at $x\to\pm\infty$. Configurations with $\varphi(+\infty) \neq \varphi(-\infty)$ are called topological, while those with $\varphi(+\infty) = \varphi(-\infty)$ are called non-topological. A configuration belonging to one topological sector can not be transformed into a configuration from another topological sector via a sequence of configurations with finite energies, i.e., through a continuous deformation.

We will denote a topological sector by two numbers --- the limits of the field at $x\to-\infty$ and $x\to+\infty$. For example, in the case given by \eqref{eq:two} the configuration belongs to the topological sector $\left(\varphi_i^{{\scriptsize\mbox{(vac)}}},\varphi_j^{{\scriptsize\mbox{(vac)}}}\right)$.

\begin{figure}[h!]
\centering
\includegraphics[scale=0.2]{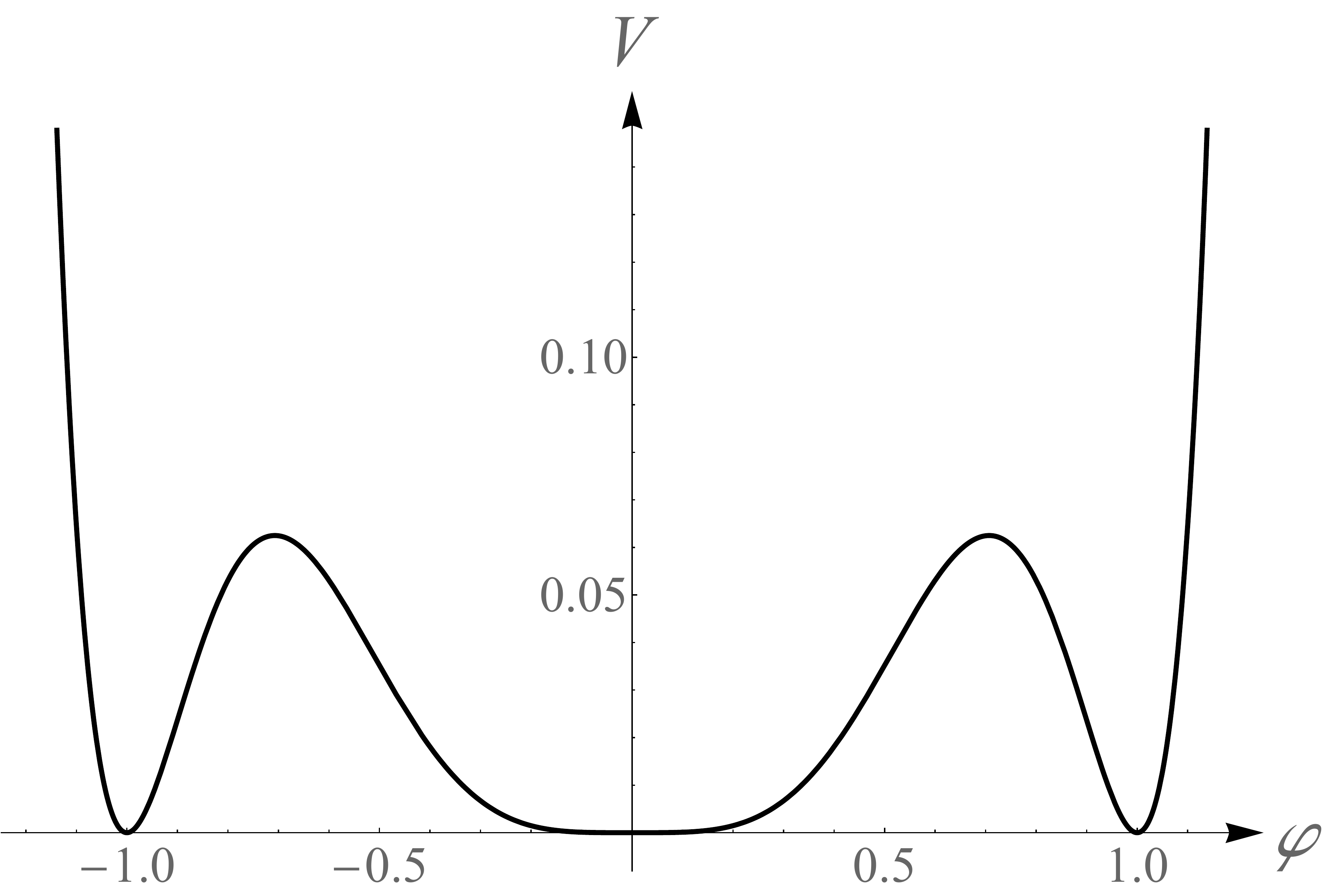}
\caption{The potential of the $\varphi^8$ model.}
\label{fig:potential}
\end{figure}

\begin{figure}[h!]
\vspace{-10ex}
\centering
\includegraphics[scale=0.4]{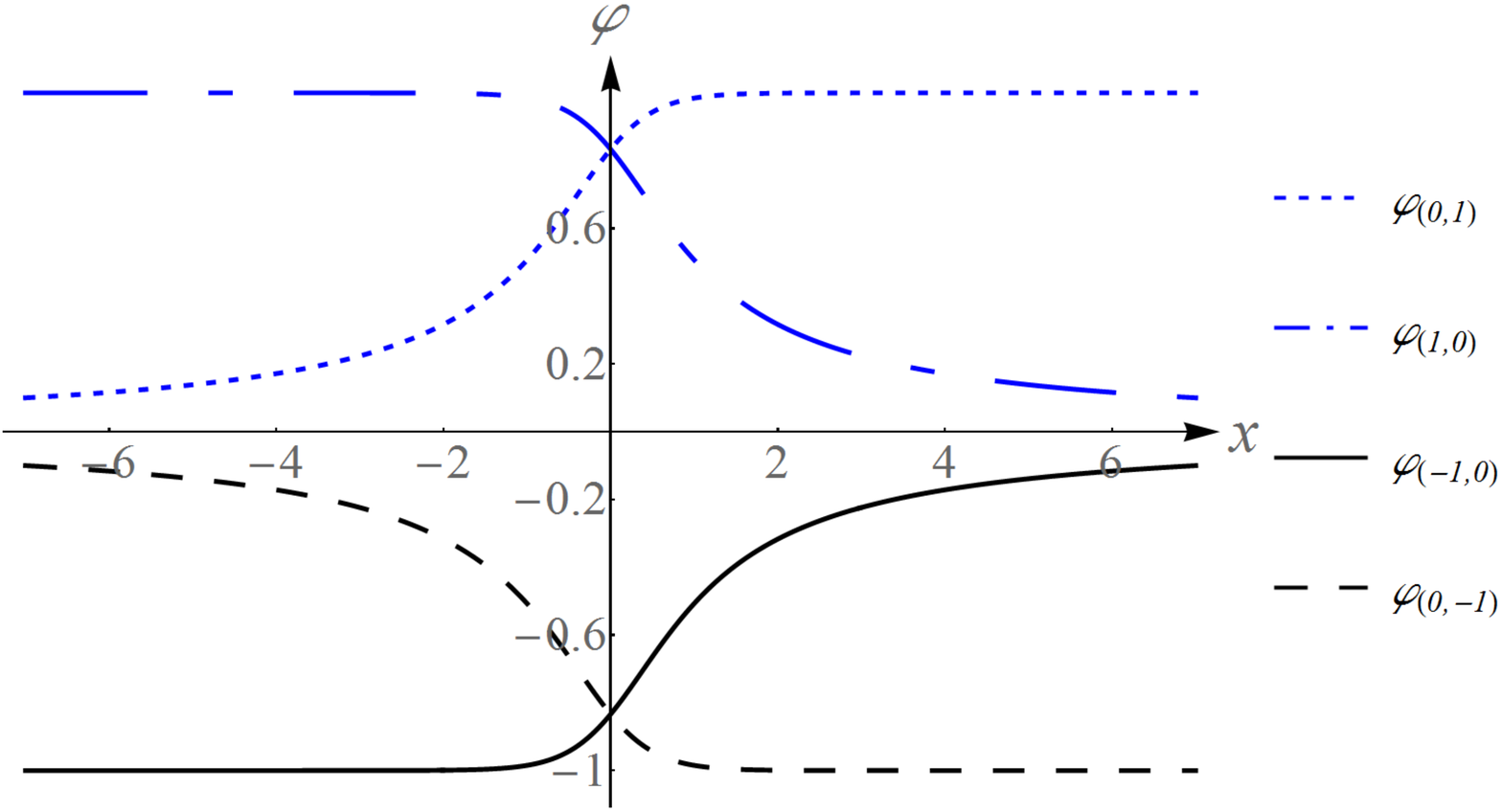}
\caption{Kinks and antikinks of the $\varphi^8$ model.}
\label{fig:kinks}
\end{figure}

The $\varphi^8$ model, which we will study below, is defined by the Lagrangian \eqref{eq:lagrangian} with the potential
\begin{equation}
\label{eq:potential}
V(\varphi)=\varphi^4(1-\varphi^2)^2.
\end{equation}
The potential \eqref{eq:potential} has three degenerate minima, $\varphi_1^{{\scriptsize\mbox{(vac)}}}=-1$, $\varphi_2^{{\scriptsize\mbox{(vac)}}}=0$, and $\varphi_3^{{\scriptsize\mbox{(vac)}}}=1$, see Fig.~\ref{fig:potential}.
The model possesses two kinks, $\varphi_{(-1,0)}(x)$ and $\varphi_{(0,1)}(x)$, and two corresponding antikinks, $\varphi_{(0,-1)}(x)$ and $\varphi_{(1,0)}(x)$. The expressions for the kinks and antikinks can be obtained only implicitly. In particular, substituting the potential \eqref{eq:potential} into Eq.~\eqref{eq:BPS} and integrating, we come to the following implicit expression for kinks $(-1,0)$ and $(0,1)$:
\begin{equation}
\label{eq:kinks}
2\sqrt{2}\: x = -\frac{2}{\varphi}+\ln\frac{1+\varphi}{1-\varphi}.
\end{equation}
The corresponding antikinks can be obtained from \eqref{eq:kinks} by the transformation $x\to-x$:
\begin{equation}
\label{eq:antikinks}
2\sqrt{2}\: x = \frac{2}{\varphi}-\ln\frac{1+\varphi}{1-\varphi}.
\end{equation}
The kinks and antikinks of the $\varphi^8$ model are shown in Fig.~\ref{fig:kinks}.
The mass of a kink can be found from Eq.~\eqref{eq:stenergy}, it is the same for all kinks and antikinks of the model:
\begin{equation}
\label{eq:kinkmass}
M_\mathrm{k} = \frac{2\sqrt{2}}{15}.
\end{equation}
A moving kink (antikink) can be obtained by the Lorentz boost, its energy depends on the velocity $v$ as
\begin{equation}
E_\mathrm{k} = \frac{M_\mathrm{k}}{\sqrt{1-v^2}}.
\end{equation}

Each kink has exponential asymptotics at one spatial infinity and power-law asymptotics at the other. For the kinks $\varphi_{(-1,0)}(x)$ and $\varphi_{(0,1)}(x)$ we have
\begin{equation}\label{eq:kink1_asymp_minus}
\varphi_{(-1,0)}(x)\approx-1+\frac{2}{e^2}\: e^{2\sqrt{2}\: x},\quad x\to -\infty,
\end{equation}
\begin{equation}\label{eq:kink1_asymp_plus}
\varphi_{(-1,0)}(x)\approx-\frac{1}{\sqrt{2}\: x},\quad x\to +\infty,
\end{equation}
\begin{equation}\label{eq:kink2_asymp_plus}
\varphi_{(0,1)}(x)\approx-\frac{1}{\sqrt{2}\: x}, \quad x\to -\infty,
\end{equation}
\begin{equation}\label{eq:kink2_asymp_minus}
\varphi_{(0,1)}(x)\approx 1-\frac{2}{e^2}\: e^{-2\sqrt{2}\: x}, \quad x\to +\infty.
\end{equation}

Below we use the following notation: the kinks belong to the sectors $(-1,0)$ and $(0,1)$, while the antikinks belong to the sectors $(1,0)$ and $(0,-1)$. Sometimes we use the term `kink' for both kinks and antikinks for brevity.

\section{Kinks scattering at low energies. Resonance phenomena}
\label{sec:Interactions}

We performed numerical simulations of the $\varphi^8$ kink-antikink scattering. To do that, we used the initial conditions in the form of the kink $(-1,0)$ and the antikink $(0,-1)$, centered at $x=-\xi$ and $x=\xi$, respectively. The kink and the antikink are moving towards each other with initial velocities $v_\mathrm{in}$. We numerically solved the partial differential equation \eqref{eq:eqmo} with the potential \eqref{eq:potential}, employing the following initial configuration (ansatz) from which one can extract the values of $\varphi(x,0)$ and $\varphi_t(x,0)$:
\begin{equation}
\label{eq:incond}
\varphi(x,t) = \varphi_{(-1,0)}\left(\frac{x+\xi-v_\mathrm{in}t}{\sqrt{1-v_\mathrm{in}^2}}\right) + \varphi_{(0,-1)}\left(\frac{x-\xi+v_\mathrm{in}t}{\sqrt{1-v_\mathrm{in}^2}}\right).
\end{equation}
In this configuration, the kink and the antikink face each other with the power-law tails, so one would hope to capture the effects of the long-range interaction between the two solitons.

In our simulation we used the initial kink-antikink half-distance $\xi=15$. The second order partial differential equation \eqref{eq:eqmo} was solved using the standard explicit finite difference scheme,
\begin{equation}
\frac{\partial^2\varphi}{\partial t^2} = \frac{\varphi_j^{k+1}-2\varphi_j^k+\varphi_j^{k-1}}{\delta t^2}, \quad \frac{\partial^2\varphi}{\partial x^2} = \frac{\varphi_{j+1}^k-2\varphi_j^k+\varphi_{j-1}^k}{\delta x^2},
\end{equation}
where $(j,k)$ number the corresponding coordinates of the grid points, $(x_j,t_k)$, on a grid with the time step $\delta t=0.008$ and the spatial step $\delta x=0.01$. To check our numerical results, we repeated selected computations with $\delta t=0.004$ and $\delta x=0.005$. We also checked whether the total energy is conserved as the grid time increases.

\begin{figure}[h!]
\begin{minipage}[h]{0.48\linewidth}
\centering\subfigure[\ Kinks' escape at $v_\mathrm{in}=0.08010$]{\includegraphics[width=\linewidth]{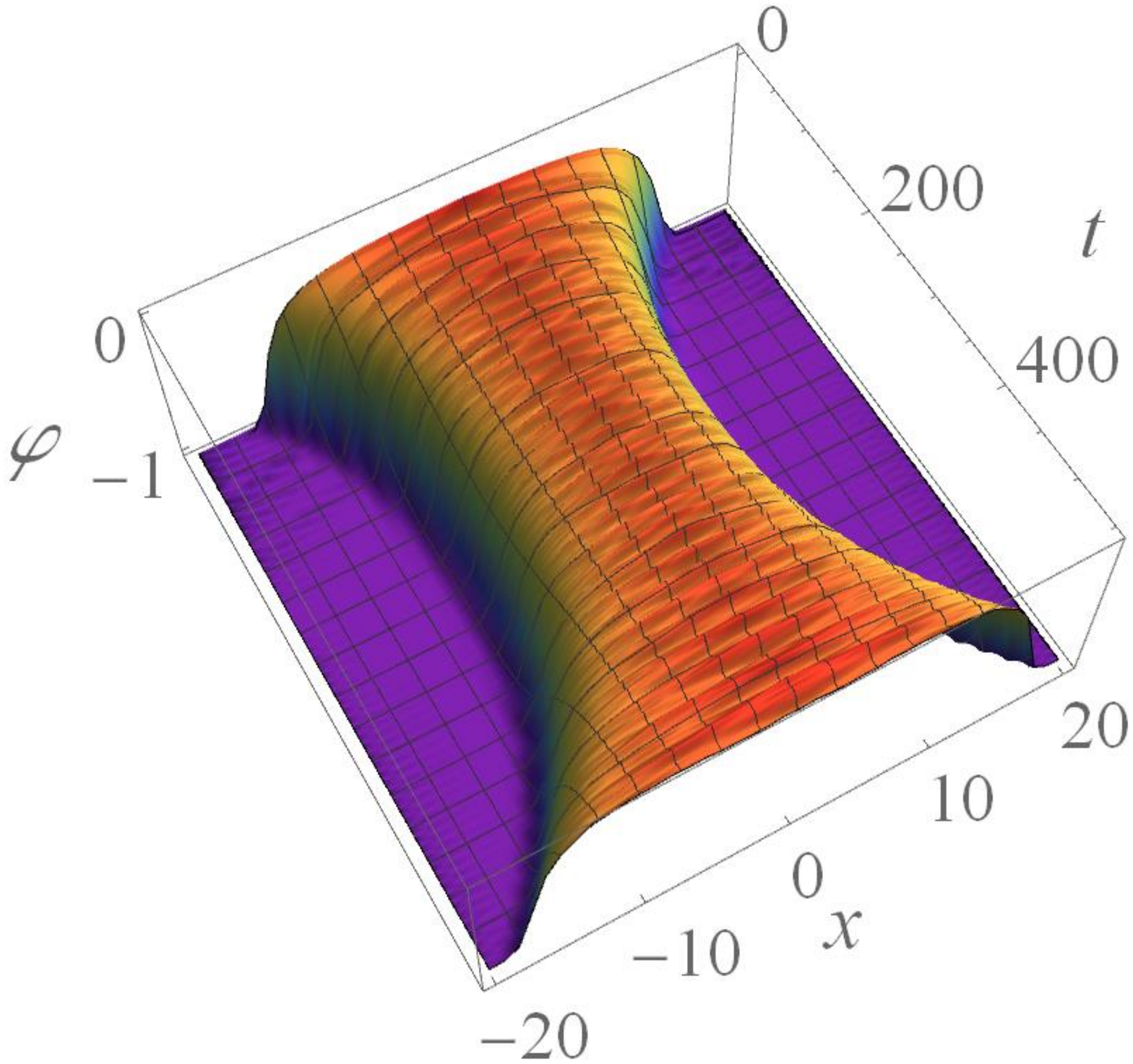}\label{fig:scat}}
\end{minipage}
\hfill
\begin{minipage}[h]{0.48\linewidth}
\centering\subfigure[\ Bounce off at $v_\mathrm{in}=0.14990$]{\includegraphics[width=\linewidth]{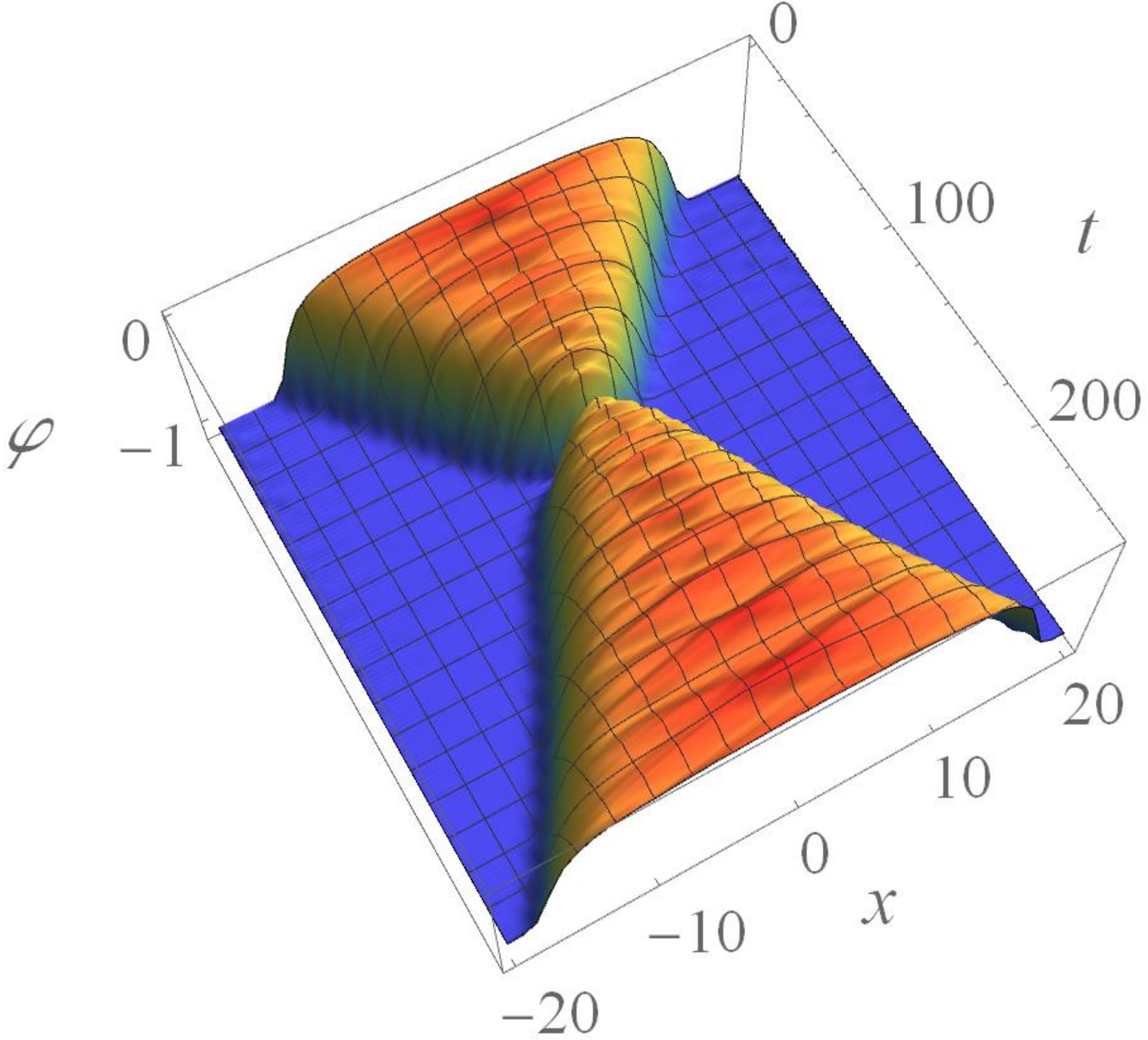}\label{fig:reflect}}
\end{minipage}
\caption{Kinks scattering at $v_\mathrm{in}<v_\mathrm{cr}^{(1)}$ (a), and at $v_\mathrm{in}>v_\mathrm{cr}^{(2)}$ (b).}
\end{figure}

The dependence of the kink-antikink scattering on the initial velocity $v_\mathrm{in}$ looks intriguing. We found two critical values of the initial velocity, $v_\mathrm{cr}^{(1)}=0.08067$ and $v_\mathrm{cr}^{(2)}=0.14791$. At $v_\mathrm{in}<v_\mathrm{cr}^{(1)}$ the kinks approaching each other stop at some distance, and then escape to infinities, see Fig.~\ref{fig:scat}. This means that kinks repel each other: if the initial velocity is too small, the kink and the antikink are not able to overcome the mutual repulsion, and do not collide. On the other hand, at initial velocities $v_\mathrm{in}>v_\mathrm{cr}^{(2)}$, the two incident kinks always escape to infinities after collision, see Fig.~\ref{fig:reflect}.

\begin{figure}[h!]
\begin{minipage}[h]{0.48\linewidth}
\centering \subfigure[\ Bion formation at $v_\mathrm{in}=0.14500$]{\includegraphics[width=1\linewidth]{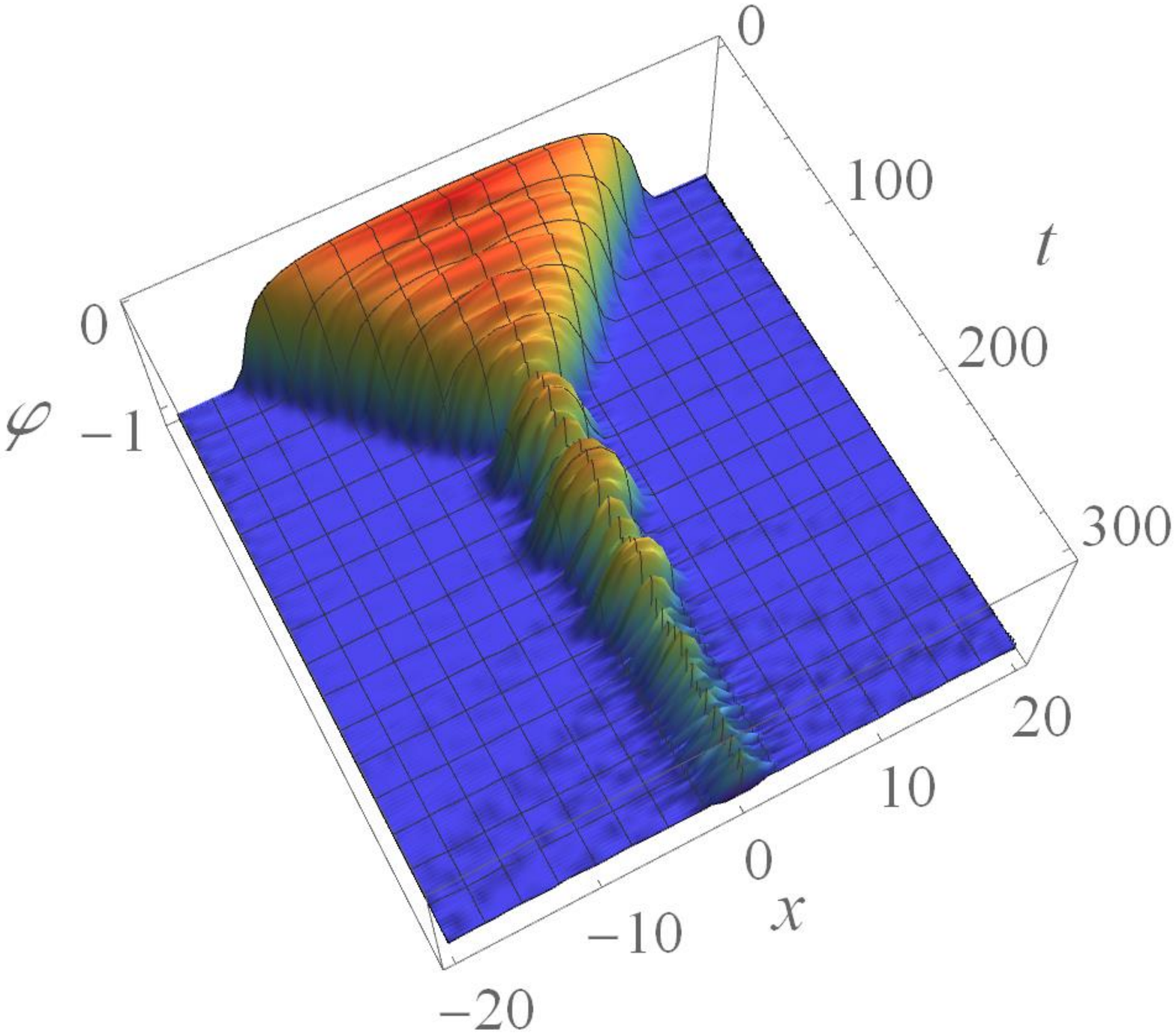} 
\label{fig:bion_3d}}
\end{minipage}
\hfill
\begin{minipage}[h]{0.48\linewidth}
\centering \subfigure[\ Two-bounce window, $v_\mathrm{in}=0.14641$]{\includegraphics[width=1\linewidth]{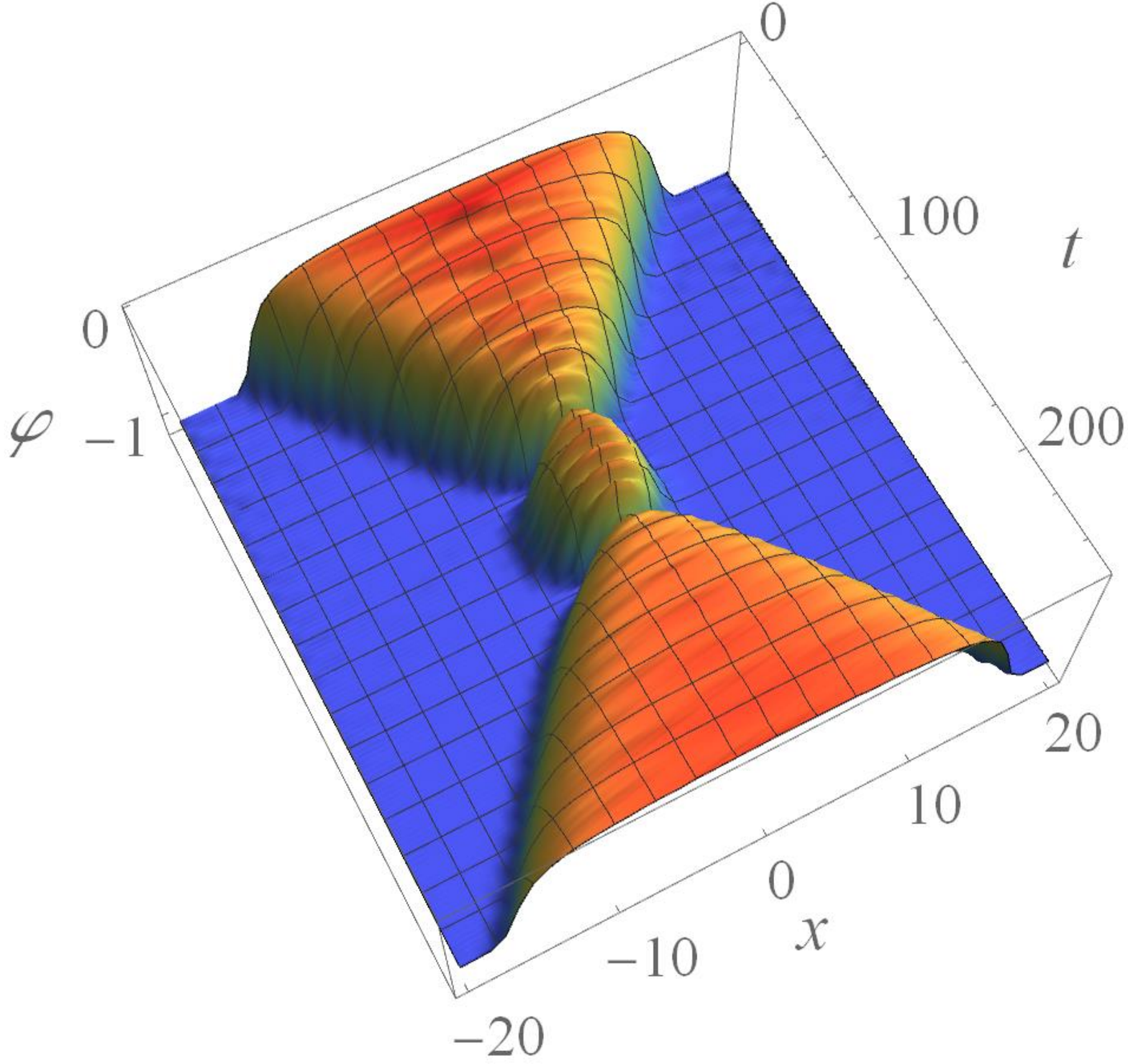}\label{fig:2win_3d}}
\end{minipage}
\begin{minipage}[h]{0.48\linewidth}
\centering \subfigure[\ Three-bounce window, $v_\mathrm{in}=0.14608$]{\includegraphics[width=1\linewidth]{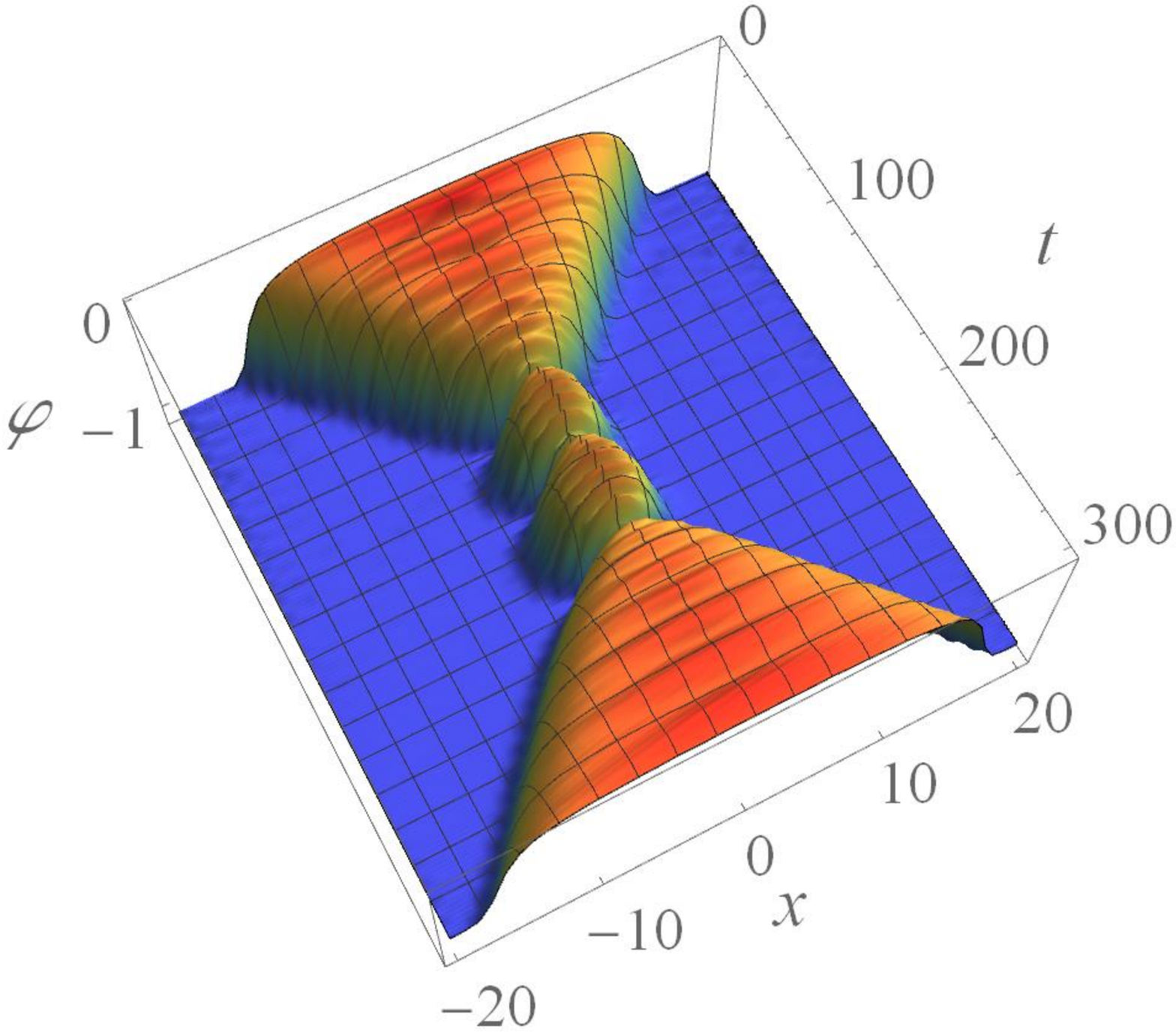}\label{fig:3win_3d}}
\end{minipage}
\hfill
\begin{minipage}[h]{0.48\linewidth}
\centering \subfigure[\ Four-bounce window, $v_\mathrm{in}=0.14726$]{\includegraphics[width=1\linewidth]{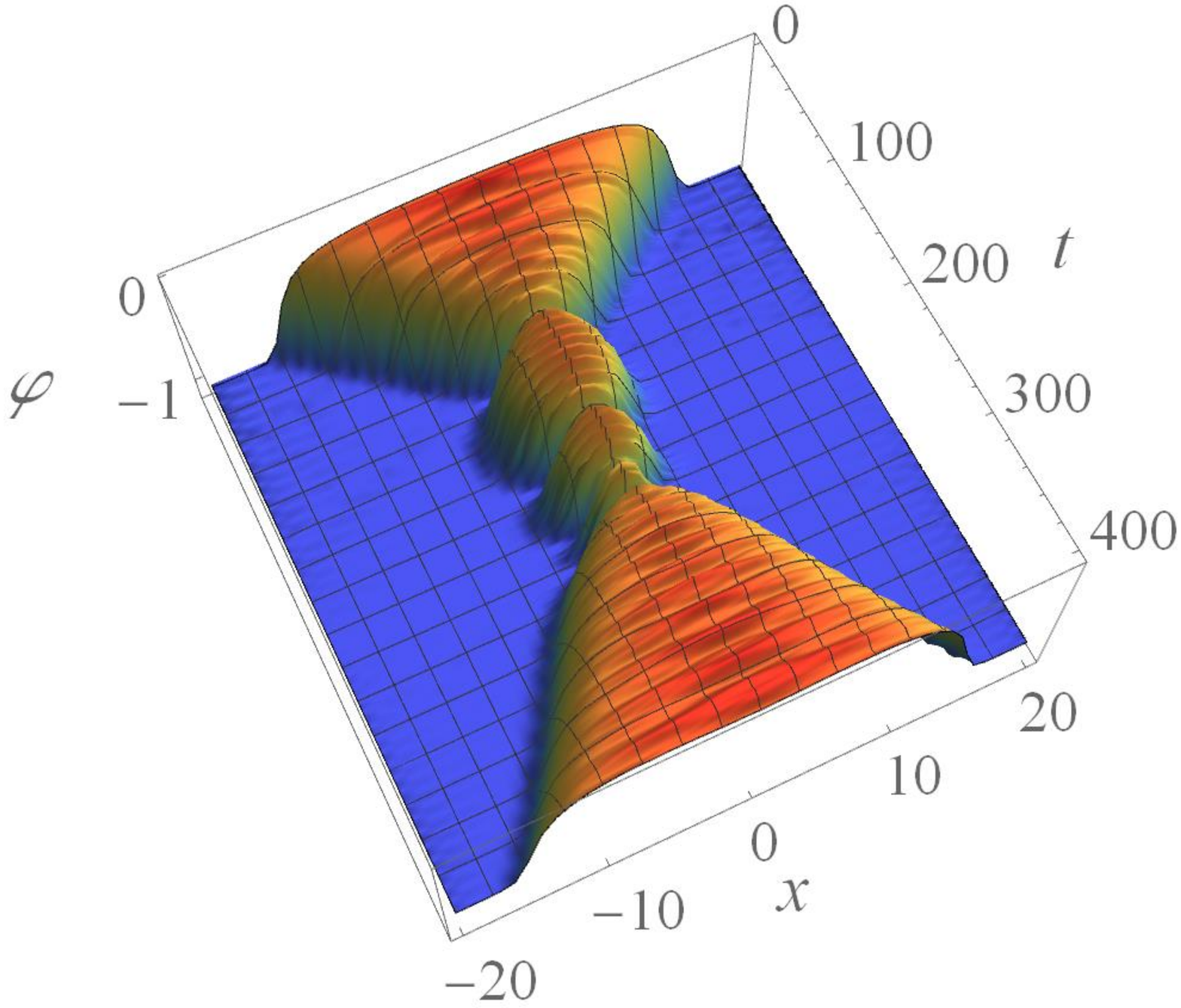} \label{fig:4win_3d}}
\end{minipage}
\caption{Kinks scattering at $v_\mathrm{cr}^{(1)}<v_\mathrm{in}<v_\mathrm{cr}^{(2)}$.}
\label{fig:space_time}
\end{figure}

The most intricate structure of the kink-antikink scattering is found in the range of initial velocities $v_\mathrm{cr}^{(1)}<v_\mathrm{in}<v_\mathrm{cr}^{(2)}$. In this regime we observe the capture and the formation of a bound state of the two kinks, see Fig.~\ref{fig:bion_3d}. There is also a complicated pattern of the so-called `escape windows' --- small ranges of the initial velocity, within which the kinks are able to escape to infinities. Note that, as opposed to the reflection at $v_\mathrm{in}>v_\mathrm{cr}^{(2)}$, within an escape window the kinks escape to infinities after two or more collisions, see Figs.~\ref{fig:2win_3d}--\ref{fig:4win_3d}.

The accepted explanation of these escape windows is the resonance energy exchange between the translational and the vibrational modes of the kinks. For example, consider a two-bounce window, i.e.\ a window within which the kinks escape after two collisions, see Fig.~\ref{fig:2win_3d}. At the first collision, a part of the kinetic energy is transferred into the vibrational modes of the kinks (or into those of the composite kink+antikink configuration). After that, due to the loss of the kinetic energy, the kinks are not able to escape, and they return and collide again. If a certain relation between the frequency of the vibrational mode and the time between the first and the second collisions holds, the energy transmitted into the vibrational mode can be returned back into the translational mode, and the kink and antikink are then able to escape.

The resonance return of the energy back to the translational mode does not need to occur after the second collision; it can also happen in any subsequent collision, so the kinks can escape to infinities after three or more collisions. According to this, the escape windows can be classed into two-bounce ones, three-bounce ones, etc., see Figs.~\ref{fig:3win_3d}, \ref{fig:4win_3d}.

\begin{figure}[h!]
\begin{minipage}[h]{0.48\linewidth}
\centering \subfigure[\ Bion formation at $v_\mathrm{in}=0.14500$]{\includegraphics[width=1\linewidth]{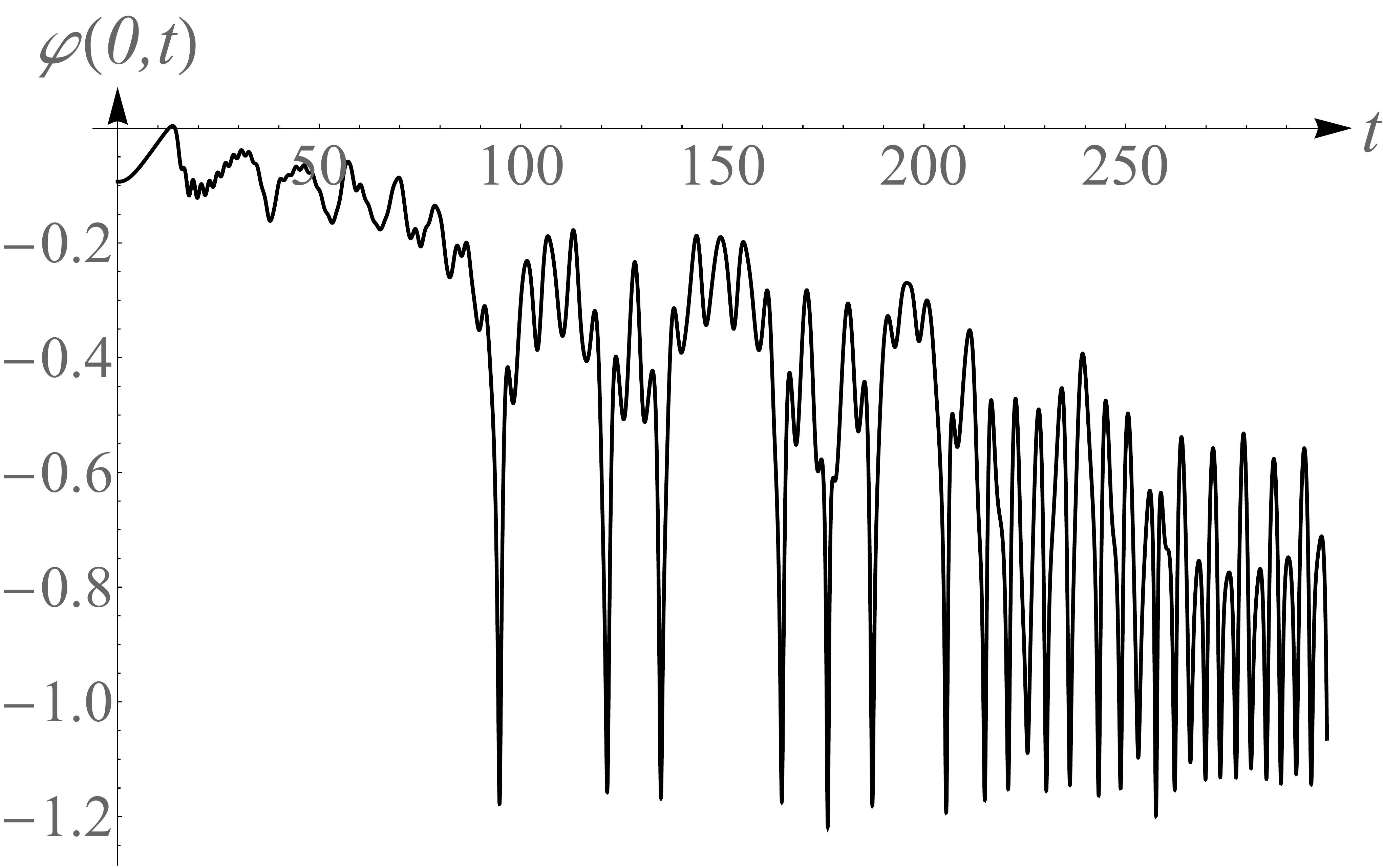} \label{fig:bion_2d}}
\end{minipage}
\hfill
\begin{minipage}[h]{0.48\linewidth}
\centering\subfigure[\ Two-bounce window, $v_\mathrm{in}=0.14641$]{\includegraphics[width=1\linewidth]{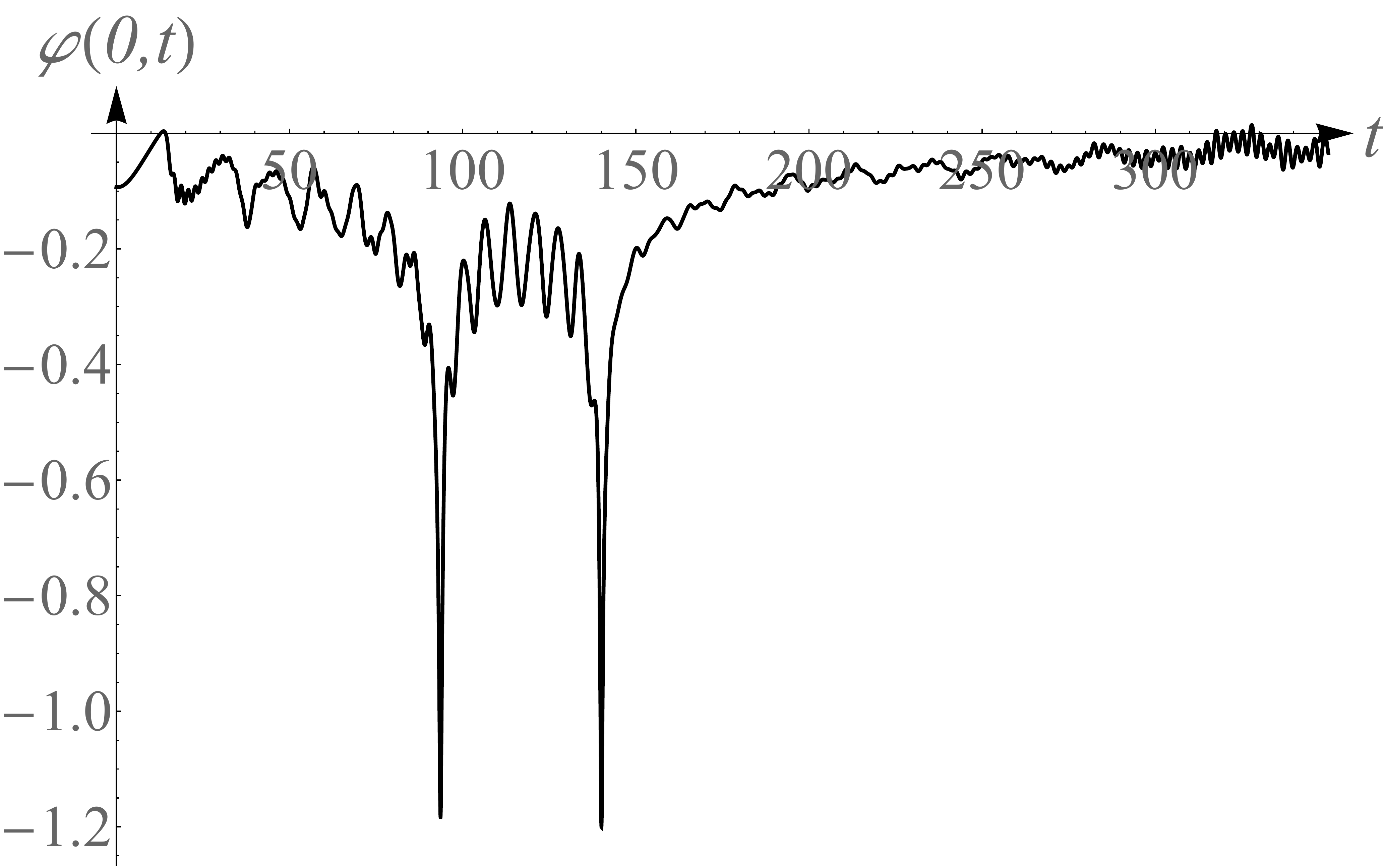} \label{fig:2win_2d}}
\end{minipage}
\begin{minipage}[h]{0.48\linewidth}
\centering\subfigure[\ Three-bounce window, $v_\mathrm{in}=0.14608$]{\includegraphics[width=1\linewidth]{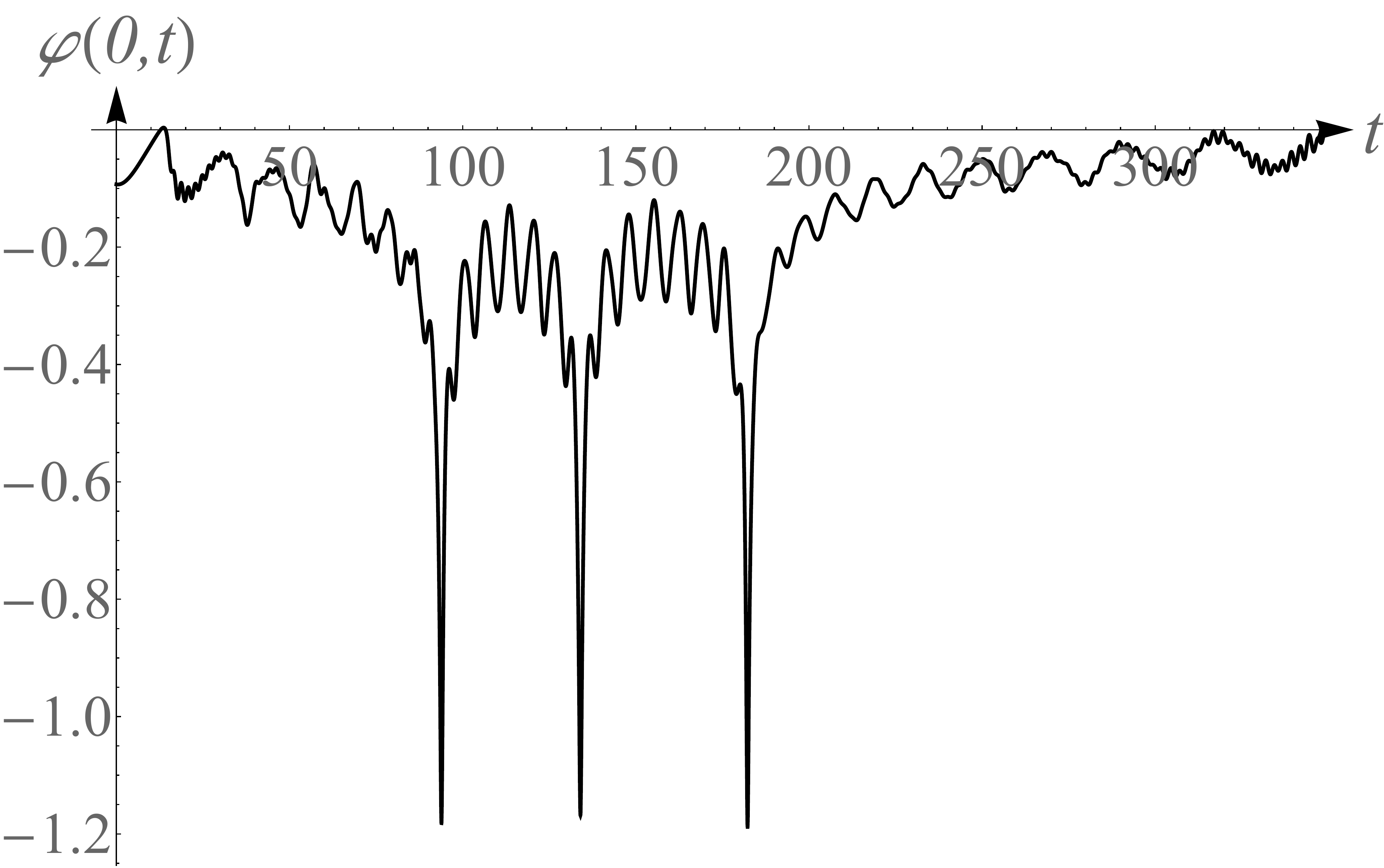} \label{fig:3win_2d}}
\end{minipage}
\hfill
\begin{minipage}[h]{0.48\linewidth}
\centering\subfigure[\ Four-bounce window, $v_\mathrm{in}=0.14726$]{\includegraphics[width=1\linewidth]{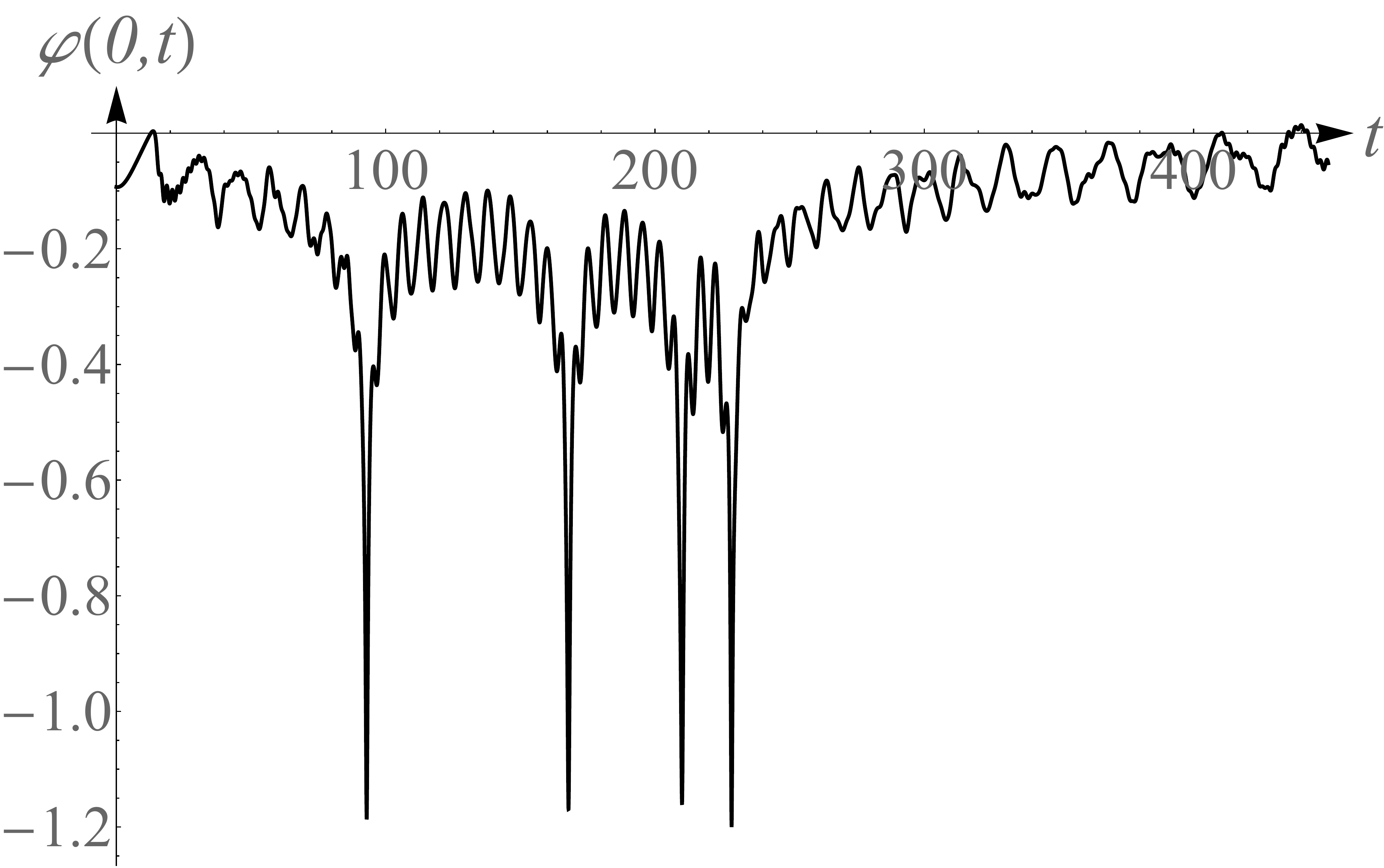} \label{fig:4win_2d}}
\end{minipage}
\caption{Time dependencies of the field at $x=0$, which correspond to the processes presented in Fig.~\ref{fig:space_time}.}
\label{fig:phi_0t}
\end{figure}

\begin{figure}[h!]
\begin{minipage}[h]{\linewidth}
\vspace{-44ex}
\centering\subfigure[\ Some of the two-bounce windows observed in the range $0.1446<v_\mathrm{in}<v_\mathrm{cr}^{(2)}$. We give $n$ --- the number of small oscillations of the field at $x=0$ between two collisions of kinks]{\includegraphics[width=1\linewidth]{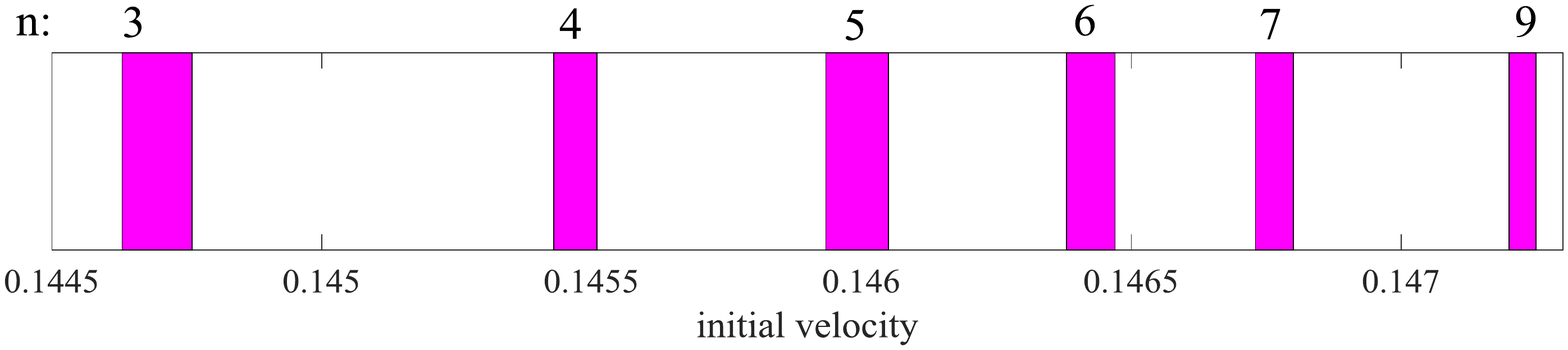}\label{fig:2_win_picture}}
\end{minipage}
\begin{minipage}[h]{\linewidth}
\vspace{-45ex}
\centering\subfigure[\ Three-bounce windows observed in the range $0.146063<v_\mathrm{in}<0.146090$]{\includegraphics[width=1\linewidth]{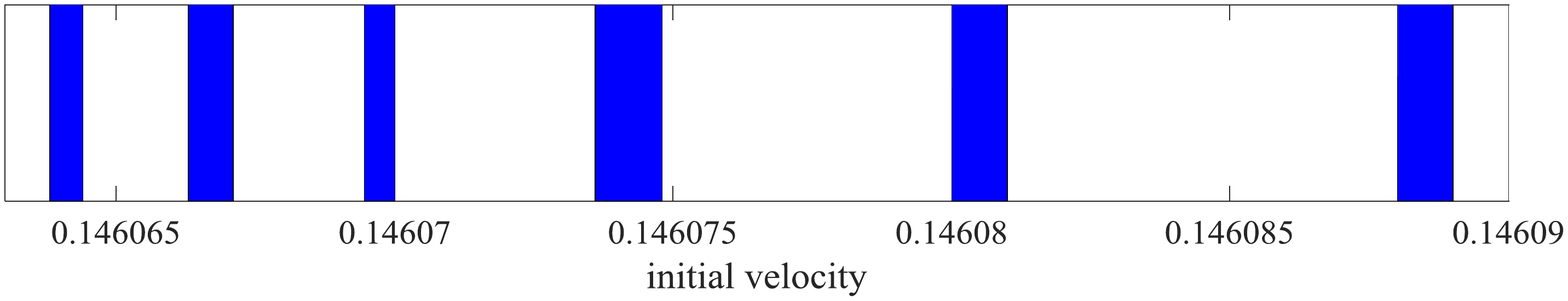}\label{fig:3_win_picture}}
\end{minipage}
\caption{Two-bounce windows (a), and one of many series of three-bounce windows (b).}
\label{fig:gen_picture}
\end{figure}

Time dependence of the field $\phi$ at $x=0$, $\varphi(0,t)$, corresponding to a bion and to two-, three-, and four-bounce windows are shown in Fig.~\ref{fig:phi_0t}. In Fig.~\ref{fig:gen_picture} we present the pattern of selected two-bounce windows and a series of three-bounce windows found by us. All escape windows are located in the range $0.1446<v_\mathrm{in}<v_\mathrm{cr}^{(2)}$ and form a fractal structure. Two-bounce windows are the broadest and located in the mentioned interval of the initial velocities, see Fig.~\ref{fig:2_win_picture}. In this figure we also give $n$ --- the number of small oscillations of the field at $x=0$ between two collisions of kinks. Near each two-bounce window, we observed a series of three-bounce windows. In Fig.~\ref{fig:3_win_picture} we show a series of three-bounce windows in the range $0.146063<v_\mathrm{in}<0.146090$. Near the three-bounce windows, we found series of four-bounce windows, and so on. We emphasize that we do not concentrate our efforts to find {\it all} existing escape windows. Our goal is to discover the phenomena qualitatively, and provide its partial quantitative investigation in order to reveal the mechanism.

As we have already mentioned, within the ansatz \eqref{eq:incond} the kink and the antikink of the $\varphi^8$ model repel each other. At the same time, at the initial velocities in the range $v_\mathrm{cr}^{(1)}<v_\mathrm{in}<v_\mathrm{cr}^{(2)}$ we observe the capture of the kinks with the formation of a bion, and a complicated pattern of escape windows. These phenomena are possible only if the incident kinks attract each other. Thus we can assume that the kink-antikink interaction has at least two regimes: at large separation the kinks repel, while being brought into contact they can become trapped. A detailed study of the kink-antikink forces within this and some other ansatzes can be a subject of another publication.

Summarizing, in the case of the kinks $(-1,0)$ and $(0,-1)$ of the $\varphi^8$ model within the ansatz \eqref{eq:incond} the repulsion leads to the appearance of the lower critical velocity $v_\mathrm{cr}^{(1)}$, so that at $v_\mathrm{in}<v_\mathrm{cr}^{(1)}$ the kinks do not collide. At $v_\mathrm{in}>v_\mathrm{cr}^{(1)}$ the $\varphi^8$ kinks behavior is rather similar to the observed previously in, e.g., the $\varphi^4$ \cite{Campbell.phi4.1983,Anninos.PRD.1991,Goodman.PRL.2007} or the $\varphi^6$ \cite{dorey,GaKuLi,MGSDJ} models. Moreover, as we show below, the situation is similar to the case of the $\varphi^6$ kinks scattering \cite{dorey}. The point is that the $\varphi^8$ kink do not possess an internal vibrational mode, but our simulations reveal a rich structure of resonance phenomena. Apparently, this is a consequence of existence of the collective vibrational modes of the composite kink+antikink configuration \cite{GaKuPRE,dorey}.

\section{Excitation spectra of the ${\boldsymbol{\varphi^8}}$ kinks and their connection with the escape windows}
\label{sec:Spectra}

This section attempts to find an explanation of the observed pattern of the escape windows. We start from a hypothesis that the resonance energy exchange between translational and vibrational modes takes place: at the initial impact a part of the kinetic energy is transferred into the vibrational energy, therefore the kink and antikink can not escape anymore, and they return and collide again. In the second collision some of the vibrational energy can be returned to the kinetic energy, allowing the kinks to escape to infinities. This is what happens within a two-bounce window. Below we analyze the excitation spectrum of an individual kink and that of a kink+antikink configuration in order to find vibrational modes that can store energy.

As we show below in this section, the $\varphi^8$ kink (antikink) does not possess an internal vibrational mode. Therefore the excitations of an individual kink can not store and return energy. At the same time, the excitation spectrum of the composite kink+antikink configuration depending on the distance between the kink and antikink, can have vibrational modes, which can accumulate energy.

In order to analyse the excitation spectrum of an individual kink, we add a small perturbation $\eta(x,t)$ to the static solution  $\varphi_\mathrm{k}(x)$, i.e., we write
\begin{equation}
\varphi(x,t) = \varphi_\mathrm{k}(x) + \eta(x,t), \quad |\eta| \ll |\varphi_\mathrm{k}|.
\end{equation}
Substituting $\varphi(x,t)$ into the equation of motion \eqref{eq:eqmo} and linearizing, we obtain:
\begin{equation}
\label{eq:eqforeta}
\frac{\partial^2\eta}{\partial t^2} - \frac{\partial^2\eta}{\partial x^2} + \left.\frac{d^2V}{d\varphi^2}\right|_{\varphi_\mathrm{k}(x)}\cdot\eta = 0.
\end{equation}
Looking for a solution of Eq.~\eqref{eq:eqforeta} in the form
\begin{equation}
\label{eq:eta}
\eta(x,t) = \chi(x)\cos\:\omega t,
\end{equation}
we come to the Schr\"odinger-like eigenvalue problem
\begin{equation}
\label{eq:Shrodinger}
\hat{H}\chi(x) = \omega^2\chi(x)
\end{equation}
with the Hamiltonian
\begin{equation}
\label{eq:Schrod_Ham}
\hat{H} = -\frac{d^2}{dx^2} + U(x),
\end{equation}
where the potential $U(x)$ is
\begin{equation}
\label{eq:Schrod_pot}
U(x) = \left.\frac{d^2V}{d\varphi^2}\right|_{\varphi_\mathrm{k}(x)}.
\end{equation}
The function $\chi(x)$ is a twice continuously differentiable and square-integrable function on the $x$-axis.

\begin{figure}[t!]
\centering
\begin{minipage}{0.47\linewidth}
\centering\includegraphics[width=\linewidth]{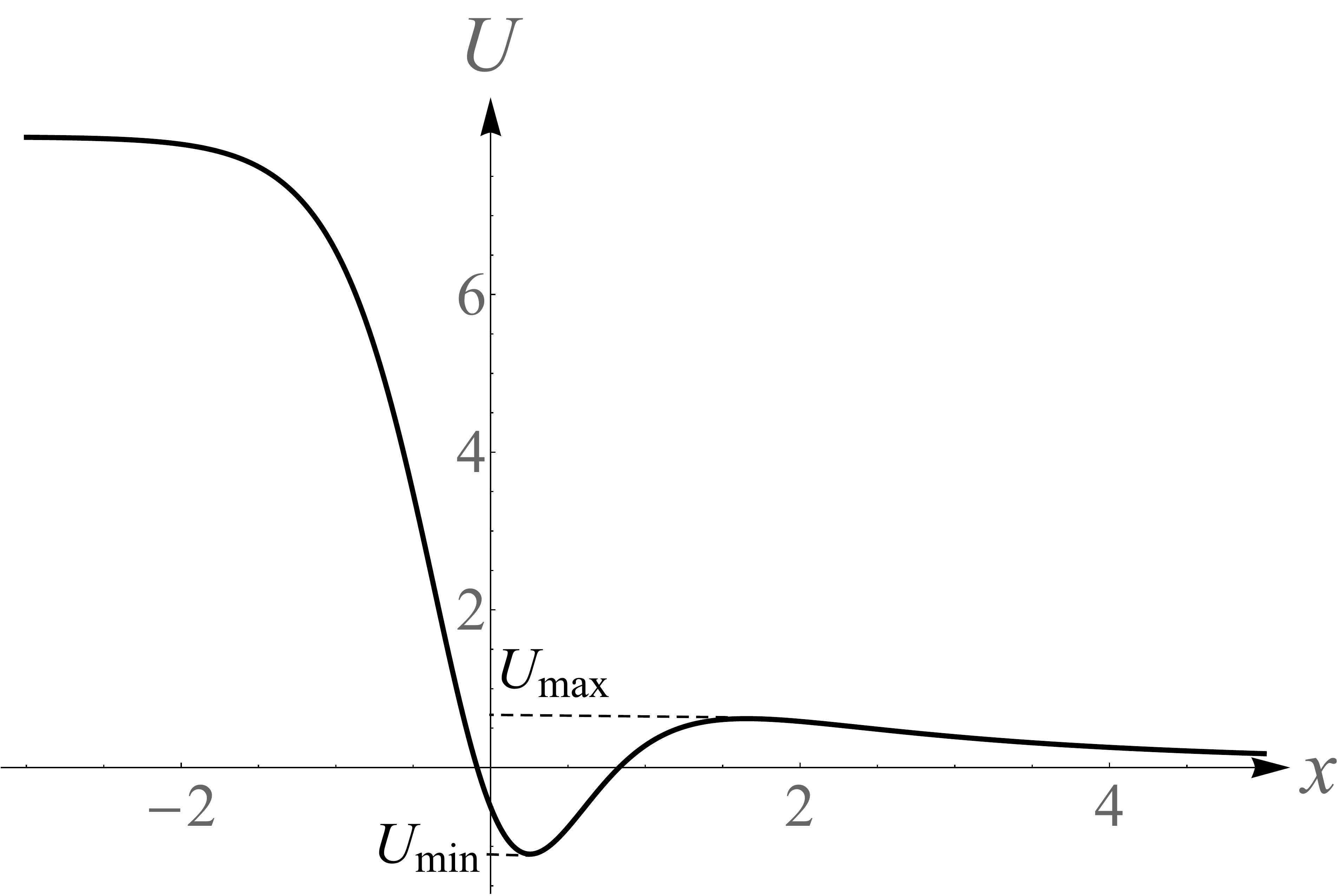}
\end{minipage}
\hspace{5mm}
\begin{minipage}{0.47\linewidth}
\centering\includegraphics[width=\linewidth]{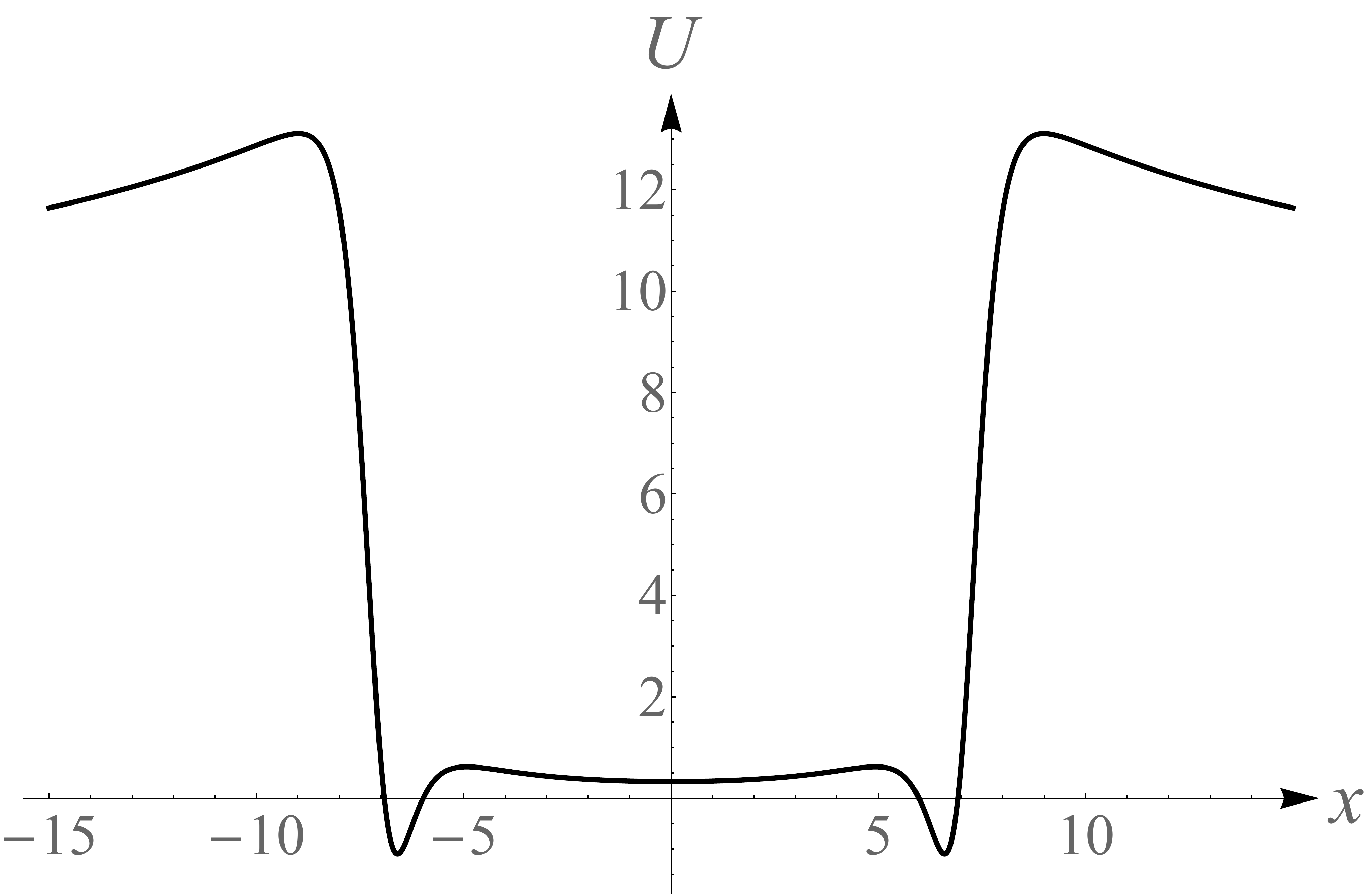}
\end{minipage}
\caption{{\bf Left panel} --- the potential \eqref{eq:Schrod_pot} for the kink (-1,0); $U(x)\to 8$ at $x\to-\infty$, $U(x)\to 0$ at $x\to+\infty$; the local minimum and the local maximum are $U_\mathrm{min}\approx -1.0985$ and $U_\mathrm{max}\approx 0.6199$, respectively. {\bf Right panel} --- the potential $U(x;\xi)$ for $\xi=7$.}
\label{fig:pot_Schr}
\end{figure}

The excitation spectra of the kink and the antikink are obviously the same. Consider the kink $(-1,0)$. The potential \eqref{eq:Schrod_pot} is presented in Fig.~\ref{fig:pot_Schr} (left panel). It can easily be shown that the discrete spectrum always has a zero (translational) mode $\omega_0=0$: taking the derivative of Eq.~\eqref{eq:steqmo} with respect to $x$, and considering that $\varphi_\mathrm{k}(x)$ is a solution of Eq.~\eqref{eq:steqmo}, we obtain:
\begin{equation}
-\frac{d^2}{dx^2}\frac{d\varphi_\mathrm{k}}{dx} + \left.\frac{d^2V}{d\varphi^2}\right|_{\varphi_\mathrm{k}(x)}\cdot\frac{d\varphi_\mathrm{k}}{dx} = 0,
\end{equation}
or
\begin{equation}
\hat{H}\cdot\frac{d\varphi_\mathrm{k}}{dx} = 0.
\end{equation}
This means that the function $\displaystyle\frac{d\varphi_\mathrm{k}}{dx}$ is an eigenfunction of the Hamiltonian \eqref{eq:Schrod_Ham} associated with the eigenvalue $\omega_0=0$. Thus a kink always has a zero mode, which simply is a consequence of the translational invariance of the Lagrangian of the
system.

Note that the zero level of the $\varphi^8$ kink's excitation spectrum is located on the upper boundary of the discrete spectrum. The zero mode wave function $\displaystyle\frac{d\varphi_\mathrm{k}}{dx}$ is nodeless and in the case of the kink $(-1,0)$ has power-law asymptotics at $x\to+\infty$. The discrete spectrum has no levels except $\omega_0=0$.

Moving on to analyze the excitation spectrum of the composite kink+antikink configuration, consider the configuration
\begin{equation}\label{eq:small_perturbation}
\varphi(x,t)=\varphi_\mathrm{k}(x+\xi)+\varphi_\mathrm{\bar{k}}(x-\xi)+\eta(x,t),
\end{equation}
where $\eta$ is a small perturbation. We consider the kink $(-1,0)$ and antikink $(0,-1)$, therefore $\varphi_\mathrm{k}(x+\xi)=\varphi_{(-1,0)}(x+\xi)$ and $\varphi_\mathrm{\bar{k}}(x-\xi)=\varphi_{(0,-1)}(x-\xi)$. Substituting Eq.~\eqref{eq:small_perturbation} into the equation of motion \eqref{eq:eqmo} and linearizing, we obtain:
\begin{equation}
\label{eq:eqforeta1}
\frac{\partial^2\eta}{\partial t^2} - \frac{\partial^2\eta}{\partial x^2} + \left.\frac{d^2V}{d\varphi^2}\right|_{\varphi_\mathrm{k}+\varphi_\mathrm{\bar{k}}}\cdot\eta = \left.\frac{dV}{d\varphi}\right|_{\varphi_\mathrm{k}} + \left.\frac{dV}{d\varphi}\right|_{\varphi_\mathrm{\bar{k}}} - \left.\frac{dV}{d\varphi}\right|_{\varphi_\mathrm{k}+\varphi_\mathrm{\bar{k}}}.
\end{equation}
Like in the case of an individual kink, substitute $\eta(x,t)$ in the form \eqref{eq:eta} into Eq.~\eqref{eq:eqforeta1} with its right-hand side set to zero. We obtain the Schr\"odinger-like eigenvalue problem:
\begin{equation}\label{eq:Shrodinger2}
\left(-\frac{d^2}{dx^2}+U(x;\xi)\right)\chi(x) = \omega^2\chi(x),
\end{equation}
where
\begin{equation}\label{eq:U_xi}
U(x;\xi)=\frac{d^2V}{d\varphi^2}\bigg|_{\varphi_\mathrm{k}+\varphi_{\bar{k}}}.
\end{equation}
The potential $U(x;\xi)$ depends on the parameter $\xi$, it is plotted in Fig.~\ref{fig:pot_Schr} (right panel) for $\xi=7$. Note that $U(x;\xi)\to 8$ at $x\to\pm\infty$ for all $\xi$.
The excitation spectrum of the composite kink+antikink configuration depends on the distance between kink and antikink, which is equal to $2\xi$.

\begin{figure}[h!]
\centering
\includegraphics[scale=0.35]{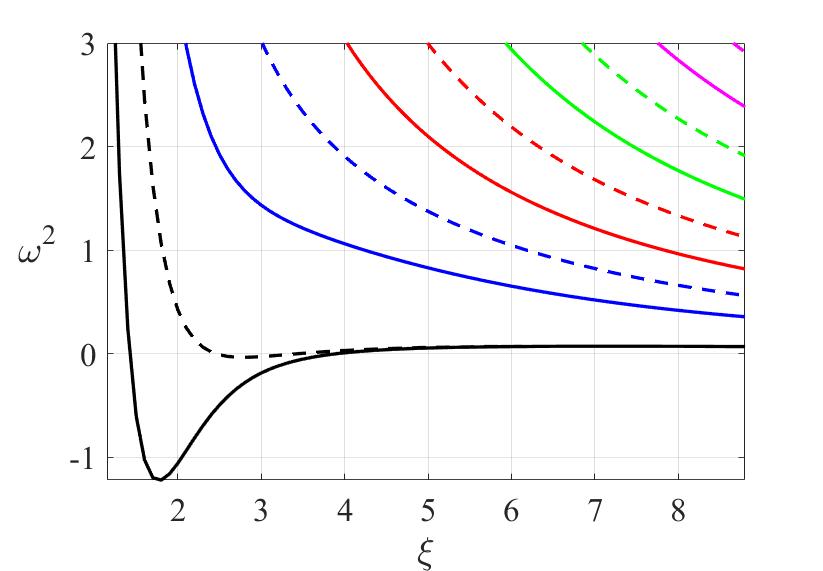}
\caption{Even (solid curves) and odd (dashed curves) bound states of the composite kink-antikink configuration.}
\label{fig:levels}
\end{figure}

We perform a numerical search of levels of the discrete spectrum of the kink+antikink configuration using a modification of the shooting method. We integrate Eq.~\eqref{eq:Shrodinger2}, using the asymptotic behavior of its solutions $\displaystyle\chi(x)\sim e^{-\sqrt{8-\omega^2}|x|}$ at $x\to\pm\infty$. We start at a large negative $x$ and obtain the `left-hand' solution, and at a large positive $x$ and obtain the `right-hand' solution. These two solutions are then matched at some point $x_0$ close to the origin $x=0$. The specific choice of $x_0$ is not very important, for example, we could take $x_0=0$. Nevertheless, a small offset from the zero value helps to avoid technical issues when $U(x;\xi)$ is an even function of $x$, and $\chi(x)$ can have a node at the origin. Eigenvalues are those values of $\omega^2$ at which the Wronskian of the `left-hand' and the `right-hand' solutions at the point $x_0$ turns to zero (changes its sign). Following this method, we obtained the curves in Fig.~\ref{fig:levels}, which we use in our analysis below.

Consider the kinks collision with the initial velocity within a two-bounce window. During the first collision of the kinks, a part of their kinetic energy is transferred into the vibrational mode(s) of the potential \eqref{eq:U_xi}, so the kinks cannot then escape to infinities, and they collide again. In the second collision, a part of the energy accumulated in the vibrational mode(s) can be returned to the kinetic energy, which results in the kinks escape to infinities. Between the two collisions the kink-antikink distance changes from zero to some maximal value $2\xi_\mathrm{max}$. However most of the time between the first and the second collisions, the kinks are separated by distance near $2\xi_\mathrm{max}$. According to this, we can assume that part of the kinetic energy is stored in a vibrational mode of the potential \eqref{eq:U_xi} with $\xi\sim\xi_\mathrm{max}$.

Note that we have retrieved the value of $\xi_\mathrm{max}$ in each collision by analyzing the space-time dependence of $\varphi(x,t)$. We also tried to apply another method, used, e.g. in Ref.~\cite{Weigel.cc.2014}. Withing this approach position of the kink can be found as
\begin{equation}
\xi(t) = \frac{\int\limits_0^\infty x\:\varepsilon(x,t)\:dx}{\int\limits_0^\infty \varepsilon(x,t)\:dx}.
\end{equation}
However in the case of kinks with power-law asymptotics this method gives big (and not constant) uncertainty in the kink position because of long tail of the second kink.


We have estimated the vibrational frequency using its obvious relation with the time $T_{12}$ between the first and the second collisions of kinks and the number $n$ of small oscillations of the field at the origin, $\omega_\mathrm{s}^{}=2\pi n/T_{12}$. Then we could estimate $\xi_{0\mathrm{s}}^{}$ using the blue solid curve (the second solid curve from the bottom) in Fig.~\ref{fig:levels}. Using this curve we assume that the first parity-even mode is excited. We have obtained $\omega_\mathrm{s}^{}=0.66\pm 0.03$ and $\xi_{0\mathrm{s}}^{}=7.8\pm 0.4$. In the numerical simulation of the kink-antikink collision at $v_\mathrm{in}=0.14790$ we observed the maximal kink-antikink separation between the bounces of the order of $\xi_\mathrm{max}\sim 6-7$. We have also performed the same analysis for some other two-bounce windows. The results are collected in Table \ref{tab:Table}.

\begin{table}
\caption{Results of the analysis of the kink-antikink collisions within some two-bounce windows. Here $\omega_\mathrm{s}^{}$ is the vibrational mode frequency obtained by the `simple' analysis; $\xi_{0\mathrm{s}}^{}$ stands for the corresponding values of the parameter $\xi$.}
\label{tab:Table}
\begin{center}
\begin{tabular}{|c|c|c|c|c|c|c|}
\hline
$n$ & $v_\mathrm{in}$ &  $\omega_\mathrm{s}^{}$ &  $\xi_{0\mathrm{s}}^{}$ & $\xi_\mathrm{max}$ \\
\hline
9 & 0.14724 &  $0.77\pm 0.09$  & $6.4\pm 1.0$ & 4--5 \\
10 & 0.14738 & $0.76\pm 0.08$  & $6.5\pm 0.9$ & 4--5 \\
11 & 0.14750 & $0.75\pm 0.05$  & $6.6\pm 0.8$ & 4--5 \\
12 & 0.14760 & $0.74\pm 0.06$  & $6.8\pm 0.8$ & 4--5 \\
13 & 0.14766 & $0.74\pm 0.06$  & $6.8\pm 0.7$ & 5--6 \\
14 & 0.14772 & $0.73\pm 0.05$  & $6.9\pm 0.7$ & 5--6 \\
15 & 0.14777 & $0.71\pm 0.05$  & $7.2\pm 0.6$ & 5--6 \\
16 & 0.14780 & $0.71\pm 0.04$  & $7.2\pm 0.6$ & 5--6 \\
17 & 0.14783 & $0.69\pm 0.04$  & $7.4\pm 0.6$ & $\sim 6$ \\
18 & 0.14785 & $0.69\pm 0.04$  & $7.5\pm 0.5$ & $\sim 6$ \\
23 & 0.14790 & $0.66\pm 0.03$  & $7.8\pm 0.4$ & 6--7 \\
\hline
\end{tabular}
\end{center}
\end{table}

Our analysis thus confirms the hypothesis that the resonance phenomena, which we observed in the collisions of the kink $(-1,0)$ and the antikink $(0,-1)$, could be caused by the energy exchange between the kinetic energy and a vibrational mode of the composite kink-antikink configuration.

\section{Oscillations about the vacua}
\label{sec:Vacua}

Unlike the well-studied $\varphi^4$ and $\varphi^6$ models, the kinks of the $\varphi^8$ model with the potential \eqref{eq:potential} have power-law asymptotic behavior. This means that the kink $(-1,0)$ substantially disturbs everything on its right, while the kink $(0,-1)$ disturbs everything on its left. In particular, in the case of the configuration $(-1,0,-1)$, the vacua $-1$ and $0$ are substantially disturbed. The vacuum $0$ between the kink and antikink is shifted down because of both kinks' tails, while the vacuum $-1$ on the left and on the right is shifted down because of power-law tail of one of the kinks. As a consequence, the initially formed static kink+antikink configuration of the type $(-1,0,-1)$ (a simple sum of the kink and antikink, Eq.~\eqref{eq:incond} with $v_\mathrm{in}=0$) is influenced by the oscillations of the field about the vacua $0$ and $-1$.

In order to analyze the oscillations about the vacuum $\varphi_i^{{\scriptsize\mbox{(vac)}}}$ of a model with a polynomial potential, we add  a small perturbation $\eta(x,t)$ to the vacuum $\varphi_i^{{\scriptsize\mbox{(vac)}}}$, i.e., we write
\begin{equation}
\varphi(x,t) = \varphi_i^{{\scriptsize\mbox{(vac)}}} + \eta(x,t),\quad |\eta(x,t)|\ll 1.
\end{equation}
Substituting $\varphi(x,t)$ into the equation of motion \eqref{eq:eqmo} and taking into account that $\varphi_i^{{\scriptsize\mbox{(vac)}}}$ is a zero of order $\nu$ of the potential $V(\varphi)$, we obtain:
\begin{equation}
\frac{\partial^2\eta}{\partial t^2} - \frac{\partial^2\eta}{\partial x^2} + \frac{1}{(\nu-1)!}\left.\frac{d^\nu V}{d\varphi^\nu}\right|_{\varphi_i^{{\scriptsize\mbox{(vac)}}}}\cdot\eta^{\nu-1} = 0.
\end{equation}
Assuming that $\eta$ does not depend on $x$, i.e.\ $\eta=\eta(t)$, we come to the ordinary differential equation
\begin{equation}
\eta^{\prime\prime} + \alpha\:\eta^{\nu-1} = 0,
\end{equation}
where $\alpha=\displaystyle\frac{1}{(\nu-1)!}\left.\frac{d^\nu V}{d\varphi^\nu}\right|_{\varphi_i^{{\scriptsize\mbox{(vac)}}}}$. Taking the initial conditions in the form $\eta(0)=\delta$, $\eta^\prime(0)=0$, we find the period of the oscillations about the vacuum $\varphi_i^{{\scriptsize\mbox{(vac)}}}$:
\begin{equation}
T_{\varphi_i^{{\scriptsize\mbox{(vac)}}}} = \sqrt{\frac{8}{\alpha\nu\delta^{\nu-2}}}\cdot\frac{\Gamma(1/\nu)\Gamma(1/2)}{\Gamma(1/\nu+1/2)},
\end{equation}
where $\Gamma(x)$ is the gamma function.

For the vacuum $\varphi_1^{{\scriptsize\mbox{(vac)}}}=-1$ of the $\varphi^8$ model with the potential \eqref{eq:potential} we have $\nu=2$, $\alpha=8$, and
\begin{equation}
T_{-1} = \frac{\pi}{\sqrt{2}}.
\end{equation}
It is interesting that the period (and frequency) of the oscillations about the vacuum $\varphi_1^{{\scriptsize\mbox{(vac)}}}=-1$ does not depend on the amplitude $\delta$.

The vacuum $\varphi_2^{{\scriptsize\mbox{(vac)}}}=0$ is a zero of the fourth order of the potential \eqref{eq:potential}, so we have $\nu=4$, $\alpha=4$, and
\begin{equation}
T_0 = \frac{\sqrt{\pi}\:\Gamma(1/4)}{\sqrt{2}\:\Gamma(3/4)}\:\frac{1}{\delta}.
\end{equation}

These oscillations about the vacua can contribute to the kink-antikink interaction via radiation pressure, and affect on the dynamics of the kink-antikink system. The oscillations in space between the kinks could lead to some radiation, which, in turn, could contribute to interaction of the kinks. Besides that, the presence of the oscillations between the kinks before the first impact could lead to the resonant energy transfer from these oscillations into the kinetic energy in the first collision.

\section{Conclusion}
\label{sec:Conclusion}

We studied the scattering of the kink $(-1,0)$ and the antikink $(0,-1)$ of the $\varphi^8$ model with the potential \eqref{eq:potential}. At this particular choice of the potential, each kink of this model has one exponential and one power-law tail. We employed the initial configuration of the type $(-1,0,-1)$, so that the kink and antikink are turned to each other by the power-law tails. We used ansatz \eqref{eq:incond}, i.e.\ simple sum of the kink and antikink.

We found that, unlike in the $\varphi^4$ and $\varphi^6$ models, the $\varphi^8$ kink $(-1,0)$ and antikink $(0,-1)$ repel each other, at least at large separations. Indeed, we observed that at the initial velocities $v_\mathrm{in}<v_\mathrm{cr}^{(1)}$ the kinks do not collide at all. For the initial separation $2\xi=30$ we found $v_\mathrm{cr}^{(1)}=0.08067$. On the other hand, at the initial velocities $v_\mathrm{in}>v_\mathrm{cr}^{(2)}=0.14791$ the kinks escape to infinities after a single impact.

The most interesting behavior was observed in the range of the initial velocities between the two critical values, $v_\mathrm{cr}^{(1)}<v_\mathrm{in}<v_\mathrm{cr}^{(2)}$. In this regime, the colliding kinks capture each other and form a bion. Moreover, we observed a fractal structure of the escape windows --- a well-known scenario, e.g., in the scattering of the $\varphi^4$ kinks. It should be noted that we have not attempted to find all the existing escape windows.

The accepted explanation of the escape windows is the resonance energy exchange between the translational and the vibrational modes of the colliding kinks. Unfortunately, the excitation spectrum of an individual $\varphi^8$ kink does not possess vibrational modes. Another way of looking at this question is to consider the resonance energy exchange between the kinetic energy and the collective excitations of the composite kink-antikink configuration. An important point is that the excitation spectrum now depends on the kink-antikink half-distance $\xi$.

To support this hypothesis, we analyzed the frequencies of small oscillations of the field at the origin between the first and the second kinks collisions in several two-bounce windows. On the other hand, we found the discrete spectrum of the composite kink-antikink configuration, depending on $\xi$. Comparing the results has shown that one of the vibrational modes is excited during the collision.

It should be noted that, on the one hand, the formation of a bion, as well as the appearance of the escape windows, indicates that the kink and the antikink attract each other. On the other hand, the presence of the first critical velocity $v_\mathrm{cr}^{(1)}$ and the accelerated motion of the kinks at large separations is an evidence of the kinks repulsion. How do the $\varphi^8$ kinks $(-1,0)$ and $(0,-1)$ interact? This study could not answer this question. It could be a subject of another detailed investigation with the use of various qualitative and quantitative methods, as well as different ansatzes. In our opinion, the kink-antikink interaction is complex, with many factors being potentially important, e.g., the power-law tails of kinks, the field oscillations about the vacuum values, etc. Notice that in Ref.~\cite{Malomed.EPL.1994} it was demonstrated that in a two-component non-linear system the following situation is possible: the full interaction potential yields repulsion and attraction forces with different spatial scales. In the case of ansatz \eqref{eq:incond} we probably have a situation of that kind.

It is important to note that the long-range interaction between the kink and antikink results in a dependence of the total energy of the configuration on the initial separation, i.e., we can not assume that within the ansatz \eqref{eq:incond} the kinks do not interact initially, at least at those initial separations which are suitable for numerical simulations. In all our numerical experiments we used the initial separation $2\xi=30$. It is evident that the change of the initial kink-antikink separation causes changes of the critical velocities $v_\mathrm{cr}^{(1)}$ and $v_\mathrm{cr}^{(2)}$, as well as alters the pattern of the escape windows. This is not the case for kinks with exponential asymptotic behavior --- for such kinks the initial separation $2\xi=30$ can be considered as infinite.

Finally, we would like to mention several issues that, in our opinion, could be a subject of future research.
\begin{itemize}
\item First, as it has already been noted, finding the force acting between the $\varphi^8$ kinks $(-1,0)$ and $(0,-1)$ is a challenge that we are faced with. The formation of a bion seems to be a kind of a threshold process. To the best of our knowledge, this is quite a new feature of the kink-antikink scattering.

Two methods could be applied in order to the estimate this force. The first one is the collective coordinate approximation with one or more degrees of freedom. The second one is Manton's method, mentioned in the Introduction. We plan to study the applicability of Manton's method in systems with kinks with power-law tails in a separate publication.

\item Second, it would be unfair not to mention that the potential \eqref{eq:Schrod_pot} has a non-trivial shape with a local maximum with $U_\mathrm{max}\approx 0.6199$, see Fig.~\ref{fig:pot_Schr} (left panel). This could lead to the appearance of quasi-bound long-living states (quasi-normal modes, see, e.g., \cite{Dorey.PLB.2018}) with energies somewhere in the range $0<\omega^2<U_\mathrm{max}$. Note that, e.g., for the two-bounce window at $v_\mathrm{in}=0.14790$ with $n=23$ we have the value $\omega_\mathrm{s}^2\approx 0.44$, which belongs to this range. Some resonance could present also above $U_\mathrm{max}$. A detailed study of the spectrum of the Schr\"odinger-like eigenvalue problem with the potential \eqref{eq:Schrod_pot} could also become an interesting part of future work.

\item Third, the numerical simulation of the evolution of the initial configuration $(0,-1,0)$, in which the kink and antikink are swapped compared to \eqref{eq:incond}, has not been included in this paper. Nevertheless, the study of such processes could also be interesting.
\end{itemize}

\section*{Acknowledgments}

The authors would like to thank Dr.~Ivan Christov, Dr.~Vadim Lensky, and Prof.~Panayotis Kevrekidis for reading the manuscript and for valuable comments. Numerical simulations were performed using resources of NRNU MEPhI high-performance computing center. This research was supported by the MEPhI Academic Excellence Project (contract No.~02.a03.21.0005, 27.08.2013).

\end{document}